\documentclass[useAMS,usenatbib]{mn2e}

\usepackage{times,epsfig,natbib,amssymb,amsmath,graphics,longtable,lscape}
\usepackage{xcolor}
\usepackage[ps2pdf]{hyperref}

\def \lsun{\ifmmode{{\rm\ L}_\odot}\else{${\rm\ L}_\odot $}\fi}

\def \msun{\ifmmode{{\rm\ M}_\odot}\else{${\rm\ M}_\odot$}\fi}

\def \rsun{\ifmmode{{\rm\ R}_\odot}\else{${\rm\ R}_\odot$}\fi}

\newcommand{\kms}{kms$^{-1}$}                         

\def \mdot{\ifmmode{{\rm\dot{M}}}\else{${\rm\dot{M}}$}\fi}

\newcommand\am{${'}$}

\newcommand\as{${''}$}

\newcommand{\R}{\textit{R}}

\newcommand{\ha}{H$\alpha${}}

\newcommand{\hii}{H\,{\sc ii}{}}

\title[SN~Ia environments]{On the environments of type Ia supernovae within
host galaxies}
\author[Anderson et al.]{J.~P. Anderson$^{1}$\thanks{E-mail:
janderso@eso.org}, P.~A. James$^{2}$, F. F\"orster$^{3,4}$, S. Gonz\'alez-Gait\'an$^{3,5}$,
\newauthor S.~M. Habergham$^{2}$, M. Hamuy$^{5,3}$, J.~D. Lyman$^{6}$\\
$^{1}$European Southern Observatory, Alonso de Cordova 3107, Vitacura, Casilla 19001, Santiago, Chile\\
$^{2}$Astrophysics Research Institute,
Liverpool John Moores University, IC2, Liverpool Science Park, 146 Brownlow Hill, Liverpool, 
L3 5RF, UK\\
$^{3}$Millennium Institute of Astrophysics,
Santiago, Chile\\
$^{4}$Center for Mathematical Modelling, Universidad de Chile, Avenida Blanco Encalada 2120 Piso 7, Santiago, Chile\\
$^{5}$Departamento de Astronom\'ia, Universidad de Chile, Casilla 36-D, 
Santiago, Chile\\
$^{6}$Department of Physics, University of Warwick, Coventry, CV
4 7AL, UK\\
}

\begin{document}

\date{}

\pagerange{\pageref{firstpage}--\pageref{lastpage}} \pubyear{2013}

\maketitle

\label{firstpage}

\begin{abstract}
We present constraints on supernovae type Ia (SNe~Ia) progenitors through an
analysis of the environments found at the explosion sites of 102 events within star-forming
host galaxies. \ha\ and $GALEX$ near-UV images are used to trace on-going
and recent star formation (SF), while broad band $B, R, J, K$ imaging
is also analysed.
Using pixel statistics we find that SNe~Ia show the lowest degree of association
with \ha\ emission of all supernova types. It is also found that they do not trace
near-UV emission. As the latter traces SF on
timescales less than 100 Myr, this rules out any extreme `prompt' delay-times 
as the dominant progenitor channel of SNe~Ia.
SNe~Ia best trace the $B$-band light distribution of their host galaxies. 
This implies that the population within star-forming galaxies is dominated by relatively young
progenitors. 
Splitting SNe by their \textit{(B-V)} colours at maximum light, `redder'
events show a higher degree of association to \hii\ regions and are 
found more centrally within hosts. We discuss possible explanations of
this result in terms of line of sight extinction and progenitor effects.
No evidence for correlations between SN stretch
and environment properties is observed.
\end{abstract}


\section{Introduction}
Type Ia supernovae (SNe~Ia henceforth) are considered to arise
from the accretion of matter onto a white dwarf (WD) star in a binary system. This
process is thought to increase the WD mass to the point where ignition 
leads to a runaway
thermonuclear explosion of the WD system (see
\citealt{wan12} and \citealt{mao14} for recent reviews). Indeed, for the
very nearby SN~Ia SN~2011fe, the primary progenitor star was directly constrained to be
a compact degenerate object for the first time (\citealt{nug11,blo12}, also
see SN~2013dy, \citealt{zhe13}).
However, many processes remain poorly constrained
meaning that we still have little idea on the exact explosion physics or the
progenitor parameter space that permits a successful explosion. These
issues are particularly pertinent given the use of SNe~Ia in many areas
of astrophysics.
SNe~Ia have been used as accurate distance indicators, which led to the
discovery of the accelerated expansion of the Universe \citep{rie98,per99}.
In addition, SNe~Ia are the main producers of iron peak elements in the Universe
and hence are particularly important for understanding chemical evolution processes (see e.g. \citealt{mat06,kob09} and
references therein).\\
\indent While the exact details of the progenitor systems remain unclear, the two most
popular are the single degenerate (SD), and double degenerate
(DD) scenarios. In the SD scenario (e.g. \citealt{whe73}), a WD accretes matter from a companion MS or
RG star, increasing its mass towards the Chandrasekhar mass, which leads to
carbon ignition and ensuing explosion. In the DD scenario (e.g. \citealt{ibe84}), a double WD
system loses angular momentum due to gravitational wave emission, leading to
coalescence and explosion. 
There also exist scenarios where the WD ignites and explodes
at masses both below (e.g. \citealt{nom82b}) and above (e.g. \citealt{hac12},
\citealt{pak12} and references therein) the Chandrasekhar mass. Indeed, \cite{sca14}
showed that a significant fraction of `normal' SNe~Ia explode with significantly
sub-Chandrasekhar ejecta masses.
Various lines of evidence have been presented, much coming in recent years, however a
preference for one specific progenitor system for the majority of SNe~Ia
events is still lacking.\\
\indent The initial classification of SNe into massive stars which
core-collapse (CC SNe), and lower mass progenitor systems which lead to
thermonuclear explosions was given credence by the observation that CC
SNe (SNe~II and SNe~Ibc) are exclusively observed to
occur in star-forming galaxies, while SNe~Ia are found to explode in
all galaxy types (e.g. \citealt{van02}). Given that elliptical galaxies are dominated by old stellar
populations, this constrains at least a fraction of the progenitors of SNe~Ia to be similarly evolved
systems. More involved studies have investigated how the SN~Ia rate changes
with redshift together with host galaxy properties such as colour, mass, star formation
rate (SFR), and specific
SFR (sSFR) (e.g. \citealt{str04,man05,for06,sul06}). These results have 
been used to constrain the delay time distribution (DTD) of SNe~Ia
(distribution of times between epoch of star formation, SF, and explosion), with
claims of `prompt' and `tardy' components (see
e.g. \citealt{sca05,man06}). Indeed the SN~Ia rate is 
consistent with the `prompt' component being proportional to the SFR,
and the `tardy' component being consistent with stellar mass of host galaxies
(see e.g. \citealt{sul06}, and more recently \citealt{smi12}).
DTDs have also been derived using the star formation history (SFH) of galaxies 
within a given SN search program \citep{mao11,mao12}.\\ 
\indent The defining characteristic that has allowed
SNe~Ia to be used as distance indicators, is
their low intrinsic peak luminosity dispersion ($\sim$0.35 mag,
\citealt{bra93}). 
This dispersion is lowered further once correlations
between luminosities and both SN decline rates and intrinsic colours are adopted
\citep{phi93,ham96,rie96}. Investigations have proceeded to find correlations between
light-curve parameters and host galaxy properties. 
\cite{ham00} (following \citealt{ham96c}) found that brighter SNe~Ia are
preferentially found within younger stellar populations. These initial results
have been confirmed (see e.g. \citealt{joh13}) and further investigated, 
with indications that brighter events
also prefer lower metallicity galaxies \citep{gal08,how09,nei09}. These 
differences would only be important for cosmological
studies if correlations remained \textit{after} SN luminosities
had been corrected for light-curve properties. Numerous recent studies 
have concentrated on searching for such correlations with Hubble
residuals. 
Indeed, evidence has now been
presented from multiple independent investigators that Hubble residuals show
correlations with host galaxy properties, in particular galaxy mass
\citep{kel10,sul10,lam10,gup11,dan11,joh13,hay12,chi13,pan14}. This could have
important consequences for the continued use of SNe~Ia as accurate
distance indicators out to higher redshifts, if these effects are not properly
accounted for. 
While determining whether stellar population age or metallicity
is the driving factor behind these correlations is somewhat difficult, several
studies have strongly argued that progenitor age is the dominant parameter,
with lower mass, higher sSFR galaxies producing younger SN~Ia
explosions \citep{chi13,joh13,rig13}.\\
\indent While many large samples of global SN~Ia host galaxy property studies have now
been
published, studies of the environments of SNe~Ia \textit{within} host galaxies,
such as those now regularly seen for CC SNe (see
e.g. \citealt{mod08,kel12,and12} and references therein), are, to-date rare in
the literature. The obvious reason for this is the significantly older
progenitor populations of SNe~Ia --and the therefore increased delay between SF and
explosion-- implies that the majority of SNe~Ia will explode far from their birth
places. This could then restrict the relevance of information that can be
extracted from such studies. However, some work in this direction does exist.\\
\indent The first approach involved investigating the radial
distribution of SNe, to see whether different SNe~Ia preferentially
occur at specific radial positions within galaxies. Given
that stellar population properties such as age, extinction, and metallicity
change with radial positions, one can associate SN properties
to those of the stellar populations found at those same regions.
\cite{wan97} first suggested that there is a deficit of SN~Ia in the central parts of galaxies,
which was not observed for CC SNe.
\cite{iva00} correlated SN properties with galactocentric radii and found 
that the range in brightness and light-curve width of SNe~Ia
increased with increasing radial distance from the centres of host galaxies,
suggesting this was a progenitor age effect. 
\cite{for08} analysed the radial distribution of SN~Ia in elliptical galaxies, 
finding that there was no statistical difference between the radial distribution of 
SN~Ia and the light profile of their hosts. They concluded that this implied 
that some SN~Ia progenitors do have delay times of several Gyr.
More recently,
\cite{gal12} found that the average SN extinction and colour was correlated with radial
position. \cite{wan13} have claimed that two
distinct populations of SNe~Ia exist, apparently implied from differences in the
ejecta velocities of SNe found within the inner and outer regions of host
galaxies. However, recently \cite{pan15} presented a separate analysis where the same
trend was observed but with less significance.\\
\indent To further investigate the local environments of SNe, one
can analyse the properties of the stellar populations at the exact site of SNe.
\cite{kel08} employed pixel statistics (in a very similar fashion to that
which we will use in the current analysis) finding that the SNe~Ia population
statistically followed the distribution of $g'$-band light of their hosts, 
in a similar fashion to SNe~II. Comparing those statistics to an
analytical model of spiral galaxy light distributions, \cite{ras09} concluded
that even the most `prompt' SNe~Ia have delay times of more than a few 100
Myrs. \cite{wan13} found that SNe~Ia with higher ejecta velocities
fall on brighter regions of their hosts than those with lower measured
velocities. In addition, \cite{rig13} showed that SNe~Ia occurring in regions with
detected \ha\ emission are `redder' than those exploding where no emission is found. Meanwhile,
\cite{gal14} published a study of SNe environments using 
integral field spectroscopy (IFS). They showed through a variety of indicators that SNe~Ia
were found to occur further from star-forming regions than CC SNe. In addition they
found that the SNe~Ia location properties were on average the same as those measured globally.
It is clear that future IFS studies will be particularly revealing for SN environment analyses.
Most recently \cite{kel14} demonstrated that SN~Ia falling on regions of high UV surface brightness
and SF density, appear to show significantly lower Hubble residuals, hence
improving the accuracy of a sub-set of SN~Ia as distance indicators.
This result was actually predicted in \cite{chi14} who suggested
that SN~Ia selected exclusively from star-forming host galaxies would yield
a more cosmologically uniform sample.\\
\indent In summary, SN environment analyses to-date have shown some intriguing results, which
despite the significant delay times of their progenitors, demonstrate that
these types of studies can indeed be revealing for SNe~Ia. 
As expected, SNe~Ia explode in distinct environments to CC SNe (see e.g. initial study
in \citealt{jam06}), confirming their older progenitors.
There appears to be a deficit of SNe in the central regions of star-forming
galaxies, which perhaps implies that the old bulge populations of these galaxies
are not significant producers of SNe~Ia. Meanwhile, environment studies
suggest that at least a significant fraction of SN~Ia colour diversity
is produced by host galaxy extinction, due to the preference of `redder'
SNe for more central regions and for higher \ha\ flux environments. 
There is also an intriguing suggestion that environments can be used to
select SNe samples with lower Hubble residuals.
However, the above also
shows that environment studies are still somewhat in their infancy,
particularly when compared to studies of
light-curve properties. 
Investigations have generally concentrated on one specific measurement,
and often suffer from insufficient statistics. Hence, a more wide ranging study of SN~Ia environments, which
attempts to bring together many of the aspects of the above is warranted.\\
\indent In the current paper we further investigate the properties of the 
immediate environments of
SNe~Ia within star-forming host galaxies, using multi-colour host galaxy
pixel statistics, and the radial distribution of
events with respect to both SF and the older stellar
continuum population. This follows from
\cite{jam06} where a small sample of SNe~Ia were analysed using these techniques.\\
\indent The paper is organised as follows. In the next section we discuss the SN and
galaxy samples, and summarise how we obtained our data and their
reduction. In \S\ 3 the statistical methods we
employ are discussed, and in \S\ 4 the
results are presented. This is followed by a discussion of
their implications for our understanding of SNe~Ia transient and
progenitor properties in \S\ 5. Finally we summarise and list our
main conclusions in \S\ 6. 

\section{Supernova and host galaxy samples}
The sample we study is a compilation of SNe and their host
galaxies from the literature. Initially, the sample was formed from SNe which
had occurred within the \ha\ Galaxy Survey (\ha GS; \citealt{jam04}), a
representative survey of star-forming galaxies within the local
Universe. Since the analysis of those data \citep{jam06}, a significantly
larger sample of \ha\ and $R$- or $r'$-band host galaxy observations have been
obtained. The only
criterion for obtaining these data was that a) host galaxies had recession
velocities less than 6000\kms\ so that we can probe distinct stellar
populations\footnote{A couple of slightly higher recession velocity
host galaxies are included, but make no difference to the overall results presented in the paper.}, 
b) that host galaxies had
major to minor axis ratios of less than 4:1 to reduce issues with chance
superpositions of foreground or background stellar populations onto SNe
explosion sites, and c) hosts were targeted where individual SNe
had published photometry, so that light-curve
parameters could be derived and compared to environmental information. 
Our initial aim to investigate the association of
SNe~Ia with SF within galaxies restricts our sample to star-forming late-type
galaxies. Later we will analyse and discuss possible systematics that this may
bring to our conclusions.
The sample of SNe and their host galaxies is listed in Table 1, together with
derived SN light-curve parameters. Before describing
host galaxy observations, we outline the methods for extracting light-curve
parameters for our sample using literature photometry.\\

\subsection{Light-curve parameters}
In addition to analysing the environments of SNe~Ia as an overall population, 
it is also fruitful to analyse whether specific SN~Ia properties show correlations
with their environment. This is motivated by differences
in SNe light-curve parameters found with global galaxy properties. 
Observables of individual SNe~Ia can be extracted from their optical
light-curves.
To proceed, literature photometric data was searched for all  
SNe in our sample. Where useful photometry was found
we list the source in Table 1. We then use
\textit{SIFTO} \citep{con08}, a light-curve fitter that uses spectral 
template time series that are adjusted to the observed colours of SN~Ia photometry 
to obtain values of $B$-band maximum date, a light-curve 
width parameter or `stretch' of the time-axis relative to the template, 
and multiplicative factors for each band from which an observed 
\textit{(B-V)}$_{max}$ colour is obtained. 
In order to include photometry in our analysis we require that for colour
estimations 
at least one photometric point is available between --15 and +8 days, and 
one available between +5 and +25 days with respect to $B$-band maximum
in both the $B$ and $V$ bands. In the case of stretch we include SNe if data is available
in only one of these bands (or other filters).
Furthermore, we only include light-curve fits which give
stretch values in the range 0.4--1.6.
Redshift and Galactic extinction are additional inputs to the fitter. 
The fitted colour is a combination of intrinsic colour and host reddening, 
for which no extinction law is applied.  
We recall that stretch is highly correlated with SN brightness, with 
higher stretch events being intrinsically brighter SNe \citep{per97}. 
The resulting light-curve parameters
are listed in Table 1.\footnote{While it would seem interesting to correlate 
environmental information with Hubble residuals, once we make a redshift cut to the
SNe with light-curve information (to remove the effect of
galaxy peculiar velocities) the resulting sample is too small to apply
such statistical tests as outlined here.}\\

\subsection{Host galaxy observations}
The principal aim of this work is to analyse the association of SNe~Ia with SF
within their host galaxies. To proceed with this aim we use our own
\ha\ imaging, plus $GALEX$ near-UV imaging taken from the $GALEX$ database\footnote{http://archive.stsci.edu/index.html}. 
The original \ha\ imaging galaxy sample was
taken from the \ha GS survey. Additional data has since been obtained through various
time allocations, and the sample we present here is of \ha\ plus $R$- or
$r'$-band (henceforth we will refer to both of these as simply `$R$-band') imaging of 
102 SN~Ia host galaxies. In addition to \ha\ and near-UV imaging which we use to trace
SF on differing time scales, we also analyse host environment properties
using $R$-band imaging together with $B$-band, plus $J$- and $K$-band near-IR images. Next each dataset which
forms our sample is discussed in detail, documenting how they were obtained, their reduction, and finally 
what type of information can be gained by analysing host galaxy imaging 
at those specific wavebands. 



\begin{table*} \centering
\caption{SN and host galaxy parameters}
\begin{tabular}[t]{ccccccc}
\hline
\hline
SN & Host galaxy &Galaxy type& V$_\textit{r}$ (\kms ) &
Stretch & \textit{(B-V)}$_{max}$& Photometry references\\
\hline
1937C  &  IC 4182   & SAm & 321  &1.227 &$\cdots$ & \cite{pie95} \\
1954B  &  NGC 5668  & SAd & 1577 &$\cdots$&$\cdots$ &$\cdots$  \\
1957A  &  NGC 2841  & SAb & 638  & $\cdots$&$\cdots$ &$\cdots$	\\
1963I  &  NGC 4178  & SBdm& 374  & $\cdots$&$\cdots$ &$\cdots$  \\
1963J  &  NGC 3913  & SAd & 954  & $\cdots$&$\cdots$ &$\cdots$   \\
1968E  &  NGC 2713  & SBab& 3922 & $\cdots$&$\cdots$ &$\cdots$  \\
1968I  &  NGC 4981  & SABbc&1680 & $\cdots$&$\cdots$ &$\cdots$   \\
1969C  &  NGC 3811  & SBcd& 3105 & $\cdots$&$\cdots$ &$\cdots$     \\
1971G\footnote{}&NGC 4165&SABa&1859& $\cdots$&$\cdots$ &$\cdots$    \\
1972H\footnote{}&NGC 3147&SAbc&2802&  $\cdots$&$\cdots$ &$\cdots$  \\
1974G  &  NGC 4414  & SAc & 716  & $\cdots$&$\cdots$ &$\cdots$      \\
1975A  &  NGC 2207  & SABbc&2741 & $\cdots$&$\cdots$ &$\cdots$      \\
1979B  &  NGC 3913  & SAd & 954  & $\cdots$&$\cdots$ &$\cdots$      \\	 
1981B  &  NGC 4536  & SABbc&1808 & $\cdots$&$\cdots$ &$\cdots$		 \\
1982B  &  NGC 2268  & SABbc&2222 & $\cdots$&$\cdots$ &$\cdots$		 \\
1983U  &  NGC 3227  & SAB & 1157 & $\cdots$&$\cdots$ &$\cdots$      \\
1986A  &  NGC 3367  & SBc & 3040 & $\cdots$&$\cdots$ &$\cdots$      \\
1986G  &  NGC 5128  & S0\footnote{}&547&0.717&0.839& \cite{phi87}  \\
1987D  &  MCG +00-32-01& SBbc& 2217& $\cdots$&$\cdots$ &$\cdots$	\\
1987O  &  MCG +02-20-09& S  & 4678 & $\cdots$&$\cdots$ &$\cdots$     \\
1989A  &  NGC 3687  & SABbc&2507 & 	 $\cdots$&$\cdots$ &$\cdots$ 	 \\
1989B  &  NGC 3627  & SABb &727  & 0.940& 0.394& \cite{wel94}  \\
1990N  &  NGC 4639  & SABbc&1018 & 1.064& 0.041& \cite{lir98} \\
1991T  &  NGC 4527  & SABbc&1736 & 1.068&0.082&\cite{for93,lir98} \\
       &            &      &     &      &     &\cite{alt04}\\
1991ak &  NGC 5378  & SBa  &3042 &  $\cdots$&$\cdots$ &$\cdots$\\
1992G  &  NGC 3294  & SAc  &1586 &  $\cdots$&$\cdots$ &$\cdots$\\
1992K  &  ESO 269-G57&SABb &3106 &	$\cdots$&$\cdots$ &$\cdots$\\
1992bc &  ESO 300-G09&Sab  &5996 &  1.059&--0.089 & \cite{ham96_2} \\
1994S  &  NGC 4495  & Sab  &4550 &  1.030&--0.021   & \cite{rie99} \\
1994ae &  NGC 3370  & SAc  &1279 &  1.054&--0.050 & \cite{alt04,rie05} \\
1995D  &  NGC 2962  & SAB0 &1966 &  1.089&0.033   & \cite{pat96,mei96} \\
       &            &      &     &      &     &\cite{alt04}\\
1995E  &  NGC 2441  & SABb& 3470 &  0.958&0.679   & \cite{rie99} \\
1995al &  NGC 3021  & SAbc &1541 &  1.074&0.107  & \cite{rie99} \\
1996Z  &  NGC 2935  & SABb &2271 & 	0.915&$\cdots$& \cite{rie99} \\
1996ai &  NGC 5005  & SABbc&946  & 	1.097&1.553   & \cite{rie99} \\
1997Y  &  NGC 4675  & SBb  & 4757&  $\cdots$&$\cdots$ &$\cdots$ \\
1997bp &  NGC 4680  & Pec  &2492 &  1.016&0.152   & \cite{jha06}\\
1997bq &  NGC 3147  & SAbc &2802 &  0.917&0.031   & \cite{jha06}\\
1997do &  UGC 3845  & SBbc &3034 &  0.975&0.018   & \cite{jha06}\\
1997dt &  NGC 7448  & SAbc &2194 &  $\cdots$&$\cdots$ &$\cdots$\\
1998D  &  NGC 5440  & Sa   &3689 &  0.869&--0.046 & \cite{jha06} \\
1998aq &  NGC 3982  & SABb &1109 &  0.986&--0.104 & \cite{rie05} \\
1998bu &  NGC 3368  & SABab& 897 &  0.974&0.269   & \cite{sun99}\\
1998dh &  NGC 7541  & SBbc &2689 &  0.939&0.088   & \cite{jha06,gan10} \\
1998eb &  NGC 1961  & SABc &3934 &   $\cdots$&$\cdots$ &$\cdots$\\
\hline	     
\end{tabular}
\end{table*}
\setcounter{table}{0}

\begin{table*} \centering
\caption{\textit{Continued...}}
\begin{tabular}[t]{ccccccccccc}
\hline
\hline
SN & Host galaxy &Galaxy type& V$_\textit{r}$ (\kms ) &
Stretch& \textit{(B-V)}$_{max}$& Photometry references\\
\hline
1999aa &  NGC 2595  & SABc & 4330&  1.113&--0.036 & \cite{alt04,jha06} \\
1999bh &  NGC 3435  & Sb  & 5158 &  0.818&0.872   & \cite{gan10} \\
1999bv &  MCG +10-25-14 & S & 5595&  $\cdots$&$\cdots$ &$\cdots$ \\
1999by &  NGC 2841  & SAb  &638  &0.618&0.484& \cite{gar04,gan10} \\
1999cl &  NGC 4501  & SAb  &2281 &0.939&1.059& \cite{kri06,jha06} \\
       &            &      &     &      &     &\cite{gan10}\\
1999cp &  NGC 5468  & SABcd&2842 &0.994&--0.027& \cite{kri00,gan10} \\
1999gd &  NGC 2623  & Pec  &5549 &0.976&0.367& \cite{jha06} \\
2000E  &  NGC 6951  & SABbc&1424 &1.052&0.130& \cite{val03,tsv06} \\
       &            &      &     &      &     &\cite{lai06}\\
2000ce &  UGC 4195  & SBb  &4888 &$\cdots$&$\cdots$ &$\cdots$\\
2001E  &  NGC 3905  & SBc  &5774 &1.002&--0.023& \cite{gan10} \\
2001ay &  IC 4423   & S    & 9067&1.600&$\cdots$& \cite{hic09} \\
2001bg &  NGC 2608  & SBb  &2135 &0.935&0.171& \cite{gan10} \\
2001cz &  NGC 4679  & SAbc &4643 &1.006&0.090& \cite{kri04} \\
2001eg &  UGC 3885  & S    &3809 & $\cdots$&$\cdots$ &$\cdots$		 \\
2002au &  UGC 5100  & SBb & 5514 & 	$\cdots$&$\cdots$ &$\cdots$	 \\
2002bs &  IC 4221   & SAc & 2889 & $\cdots$&$\cdots$ & $\cdots$ 	 \\
2002cr &  NGC 5468  & SABcd&2842 &0.945&--0.013& \cite{hic09} \\
2002er &  UGC 10743 & Sa & 2569  &0.930&0.138& \cite{pig04,gan10} \\
2002fk &  NGC 1309  & SAbc& 2136 &1.010&--0.093& \cite{hic09} \\
2003cg &  NGC 3169  & SAa & 1238 &0.984&1.110& \cite{eli06,hic09} \\
       &            &      &     &      &     &\cite{gan10}\\
2003cp &  MCG +10-12-78 & Sb&5927& $\cdots$&$\cdots$ & $\cdots$ \\
2003du &  UGC 9391  & SBdm& 1914 &1.019&--0.084& \cite{leo05_2} \\
2004bc &  NGC 3465  & Sab & 7221 &  $\cdots$&$\cdots$ & $\cdots$ \\
2004bd &  NGC 3786  & SABa& 2678 &	$\cdots$&$\cdots$ & $\cdots$ \\
2005A  &  NGC 958   & SBc & 5738 &0.977&0.986& \cite{con10} \\
2005F  &  MCG +02-23-27& S & 8545&  $\cdots$&$\cdots$ & $\cdots$ \\
2005G  &  UGC 8690  & Scd & 6938 &  $\cdots$&$\cdots$ & $\cdots$ \\
2005M  &  NGC 2930  & S   & 7382 &	1.114	&0.312& \cite{str11} \\
2005W  &  NGC 691   & SAbc& 2665 &0.954&0.144& \cite{str11} \\
2005am &  NGC 2811  & SBa & 2368 &0.778&0.056& \cite{con10} \\
2005bc &  NGC 5698  & SBd & 3679 &0.830&0.377& \cite{gan10} \\
2005bo &  NGC 4708  & SAab& 4166 &0.852&0.243& \cite{gan10,con10} \\
2005cf &  MCG -01-39-03 & S0\footnote{}&1937&0.995&--0.015& \cite{wan09_2}  \\ 
2005el &  NGC 1819  & SB0\footnote{}&4470&0.886&--0.088& \cite{con10}    \\
2005ke &  NGC 1371  & SABa& 1463 & 	0.677&0.653& \cite{hic09,str11} \\
2006D  &  MCG -01-33-34&SABab&2556& 0.818&0.105& \cite{str11} \\
2006N  &  MCG +11-08-12 &?&4280  &0.795&0.027& \cite{hic09} \\
2006X  &  NGC 4321  & SABbc&1571 &0.995&1.196& \cite{str11} \\
2006ax &  NGC 3663  & SAbc& 5018&	0.984	&--0.089& \cite{hic09,str11} \\
2006ce &  NGC 908   & SAc & 1509 & $\cdots$&$\cdots$ & $\cdots$ \\
2006mq &  ESO 494-G26&SABb& 968  & $\cdots$&$\cdots$ & $\cdots$\\
2006ou &  UGC 6588  & Sbc & 4047 &	1.393&0.345& \cite{hic12} \\
2007N  &  MCG -01-33-12&SAa& 3861&	0.524&0.988& \cite{hic09,str11} \\
\hline	     
\end{tabular}
\end{table*}
\setcounter{table}{0}

\begin{table*} \centering
\caption{\textit{Continued...}}
\begin{tabular}[t]{ccccccccccc}
\hline
\hline
SN & Host galaxy &Galaxy type& V$_\textit{r}$ (\kms ) & Stretch & \textit{(B-V)}$_{max}$& References \\
\hline
2007S  &  UGC 5378  & Sb  & 4161 &	1.095&0.116& \cite{hic09,str11} \\
2007af &  NGC 5584  & SABcd &1638&	0.953&0.073&  \cite{hic09,str11}\\
2007bm &  NGC 3672  & SAc & 1862 &	0.922&0.474&  \cite{hic09,str11}\\
2008bi &  NGC 2618  & SAab& 4031 &	$\cdots$&$\cdots$ & $\cdots$ \\
2008fv &  NGC 3147  & SAbc& 2802 &	1.104&0.165& \cite{tsv10,bis12} \\
2009ag &  ESO 492-G02& SAb& 2590 &	0.978&0.124& Unpublished, Carnegie \\
		&			&				&			&&& Supernova Project\\
2009ds & NGC 3905  & SBc & 5774  &	1.144&0.073& \cite{hic12} \\
2009ig & NGC 1015  & SBa & 2629  &	1.075&0.140& \cite{fol12,hic12} \\
2010eb & NGC 488    & SAb & 2272 &  $\cdots$&$\cdots$ & $\cdots$   \\
2011B  & NGC 2655   & SAB0/a&1400&  $\cdots$&$\cdots$ & $\cdots$   \\
2011ao & IC 2973    & SBd & 3210 &  $\cdots$&$\cdots$ & $\cdots$   \\
2011ek & NGC 918    & SABc& 1507 &  $\cdots$&$\cdots$ & $\cdots$   \\
2011dm & UGC 11861  & SABdm &481 &  $\cdots$&$\cdots$ & $\cdots$  \\
2011dx & NGC 1376   & SAcd& 4153 &  $\cdots$&$\cdots$ & $\cdots$   \\
\hline
\hline
\end{tabular}
\setcounter{table}{0}
\caption{Information on the SN sample analysed in this work,
together with estimated light-curve parameters and references for photometry. In the first
column we list the SN name, followed by the host galaxy in column 2. Then in
columns 3 and 4 respectively we list the Hubble type and recession velocity of
each host galaxy (information taken from NED: http://ned.ipac.caltech.edu/). This is followed by the derived
light-curve parameters for each SN: stretch
followed by \textit{(B-V)}$_{max}$ (the $B$-$V$ colour at maximum light). 
In
column 7 we list the reference for SN photometric data (where we only 
include data if the light-curve fits passed our selection criteria). 
\newline$^{1}$Classification from IAU SN catalogue
\newline$^{2}$Classification as SN~Ia from \protect\cite{san93}
\newline$^{3}$While this galaxy is classed as Hubble type S0 (i.e. no, or very
little SF), there is definitely large amounts of detected \ha\ line emission
produced by on-going SF
\newline$^{4}$Galaxy is classified as S0, but significant on-going SF is detected
\newline$^{5}$Galaxy is classified as S0, but significant on-going SF is
detected
}
\end{table*}

\subsection{Host galaxy \ha\ and $R$-band imaging}
\ha\ line emission within galaxies traces regions of young on-going
SF. The emission is produced from the recombination of interstellar medium
(ISM) hydrogen atoms after
ionisation from the intense UV flux of young massive stars. The massive stars which
make the dominant contribution to this ionising flux are of $>$15-20\msun, and
therefore \hii\ regions
within galaxies are thought to trace SF of ages of less than $\sim$10
Myrs \citep{ken98}. This is obviously much younger than even
the most `prompt' SN~Ia scenarios \citep{aub08}. However, it will be shown that an
investigation into the association of SNe~Ia with these regions can still
produce interesting results.\\
\indent The narrow band \ha\ imaging technique involves observing through a narrow
\ha\ filter (on the order of 50-100 \AA) together with separate broad-band
imaging to remove the continuum. Here, we use $R$ or $r'$ filters to remove
this component. 
This also allows us to use these images for further
analysis.
The initial data used for this project (imaging from \ha GS, \citealt{jam04}) were obtained with the 1.0 metre Jacobus Kapteyn Telescope (JKT), with a 
pixel scale of 0.333\as\ per pixel. In the
first paper of the current series \citep{jam06}, data and analysis were published for 12 SN~Ia host
environments. Here we significantly increase this sample size.
\ha\ and $R$-band imaging data were obtained with the Wide Field Camera (0.333\as\ per pixel image scale) 
mounted on the Isaac Newton
Telescope (INT) over the course of two observing runs during February of both 2007 and
2008. In addition, we searched through a series of
previous INT data samples taken for other projects by co-authors of the current
paper for further host galaxy imaging. During various time allocations 
between 2005 and 2009 host galaxy \ha\ and $r'$ band imaging was obtained
with RATCAM on the Liverpool Telescope (LT, \citealt{ste04}), with a pixel scale of 0.278\as\ per pixel.
Most recently \ha\ and $R$-band data were obtained 
with the MPG/ESO 2.2m telescope together with the Wide Field
Imager (WFI, \citealt{baa99}) during February 2010. WFI has a pixel scale of 0.238\as\ per pixel.\\
\indent While the exact details of observing strategy changed slightly between
different telescopes and instruments, overall procedures were very
similar. Generally three 300 second \ha\ exposures were obtained, followed by one 300 second
broad-band $R$-band images. 
Standard imaging reduction techniques were employed using IRAF\footnote{IRAF is distributed 
by the National Optical Astronomy Observatory, which is operated by the 
Association of Universities for Research in Astronomy (AURA) under 
cooperative agreement with the National Science Foundation.} to first bias-subtract
then flat-field all science frames. A range of foreground stars 
were then used to obtain a scaling factor between the narrow and
broad-band observations. After scaling, the broad-band images were 
subtracted from the \ha\ observations. This leaves
only the \ha\ line emission within host galaxies, ready for analysis. Finally,
images were astrometrically calibrated using the \textit{Starlink} package ASTROM,
enabling sub-arcsecond location of the SN explosion sites.\\

\subsection{$GALEX$ data}
\ha\ line emission within galaxies traces the most recent,
on-going SF regions of ages less than 10 Myrs. Given that
this is a much shorter timescale than even the youngest hypothesised SN~Ia
progenitors, we also choose to investigate the association of SN~Ia
explosion sites with SF of older ages. Near-UV emission within galaxies is
thought to trace SF out to older ages ($\sim$100 Myr, see e.g. \citealt{gog09}). In the rest of the manuscript
we will refer to near-UV imaging as tracing recent SF (in place of 
on-going with respect to that traced by \ha). We searched the $GALEX$ 
archive for near-UV images of our sample. Data were available for 74 SN~Ia host galaxies and
$GALEX$ near-UV `intensity maps' were obtained, which come astrometrically 
calibrated ready for analysis. These data have scales of 1\as\ per pixel.\\

\subsection{INT $B$-band data}
$B$-band imaging was obtained with the WFC on the Isaac Newton Telescope (INT). Exposure times of $\sim$500 seconds were
employed and the WFC gives images of 0.333\as\ per pixel scale. 
Standard imaging reduction techniques were employed to first bias-subtract
then flat-field all science frames.
Images were then astrometrically calibrated 
and accurate SN positions on reduced images were obtained.\\

\subsection{NOT nearIR data}
To obtain near-IR $J$- and $K$-band imaging the Nordic Optical Telescope (NOT) was employed using
NOTCam. 
The instrument was used in wide field imaging mode which results in images with
a FOV of 4\am $\times$4\am\ and a pixel scale of 0.234\as\ per pixel.
A total $J$-band 
exposure time of 200 seconds was used, split into 4$\times$50 second exposures. To account for the varying
near-IR sky, frames offset from host galaxies were observed with the same exposure times
as those used for science frames. Therefore 50 second exposures were
taken on target, then offsets of 250\as\ were applied 
to slew to sky positions. 
Within this sequence positions were dithered by 10\as\ to 
remove stars from the sky frames
and bad pixel defects on the detector. A similar process was employed for the $K$-band images
with a total exposure time of 900 seconds split into 9$\times$100 second exposures. 
Differential
flat fields were obtained at either the start or the end of each observing night.
Sky flats with low count levels were subtracted from similar frames with high 
levels and then the resulting flat field
was normalised. Each subsequent sky frame was subtracted from its corresponding science frame. Each 
resulting sky-subtracted science frame was then divided by the normalised flat field. These resulting 
frames were then combined using median stacking. 
Images were then astrometrically calibrated as above.
However, in many cases there were insufficient stars within the relatively small field
of view of NOTCam to enable an astrometric solution. In these cases SN positions were calculated
using offsets of SN coordinates from galaxy centres (with SN positions being taken from
either the IAU SN list or NED, and galaxy centre coordinates taken from NED). Offsets
were transformed to NOTCam pixel x and y offsets, and SN positions on the images were calculated using the
peak of the $J$- or $K$-band flux on each image as the galaxy centre. To check the validity of this
process, SN positions were also calculated in this way for images where an
astrometric calibration was possible and the two methods were compared. In general
SN pixel positions were consistent between the two methods, and NCR values (see below)
were also consistent between the two pixel coordinate values. The difference in
the mean NCR values between the 21 SNe where the above comparison was possible in the
$J$-band is 0.008, while for 20 SNe in the $K$-band the difference is 0.023.\\

\subsection{Stellar populations traced by different wavebands}
In this work we analyse the association of SNe~Ia within
star-forming hosts with galaxy light distribution in a range from
\ha\ through optical wave-bands to $K$-band near-IR observations. Each 
waveband analysed (\ha, near-UV, $B$-, $R$-, $J$- and $K$-band) 
traces different stellar population properties. 
\ha\ emission traces SF
on the shortest timescales of less than $\sim$10 Myr. Near-UV emission
traces SF on longer timescales out to $\sim$100 Myr. When one
moves to analyse broad-band observations at longer wavelengths direct age constraints
become less clear. However, stellar population models appear to be
in agreement that as one moves redwards then the older stellar mass
of a stellar population/galaxy starts to contribute more significantly (see e.g. \citealt{wor94}, and
\citealt{sch07} who used the models of \citealt{mar05}). While it is generally
assumed that young populations still dominate the flux detected through the $B$-band filter,
the $K$-band light in most stellar populations is assumed
to be dominated by the old stellar mass of that population (e.g. \citealt{man05}).\footnote{While
\cite{kan13} claimed that the $K$-band light in their galaxy sample may be tracing recent
SF, this was due to their sample being of infrared-bright starburst galaxies, 
significantly distinct from the current sample.}
Following the above, we can speculate that if a SN distribution better traces 
that flux observed through a redder filter, then the dominant age of the progenitor 
population may be larger than that of a distribution which traces flux observed through a bluer 
filter. The above is obviously only a first order approximation, and one
should be careful to draw strong conclusions following these arguments, especially
given the effects of metallicity and extinction. However,
in the following sections we will use this argument as a guide to understand the 
distributions derived, while stressing the uncertainties involved. In \S\ 5.1.1
we present further constraints on the stellar population properties probed by our broad-band images.\\

\section{Analysis methods}
The procedures for obtaining results from the statistical methods we use here,
were first presented in \cite{jam06}, and have been further outlined in their
application to CC SN host galaxy samples in \cite{and08},
\cite{and09,hab10,and12,hab12} and \cite{hab14}. Below we briefly outline 
our methods, but refer the reader to those previous publications for more
detailed documentation.

\subsection{Pixel statistics}
In \cite{jam06}, a pixel statistics method was introduced which was created to
give a quantitative measure of the association of individual SNe to the
underlying SF present within galaxies. This method produces, for
each pixel within an image a value (dubbed `NCR', \citealt{and08}) between 0
and 1 which 
indicates where within the overall host
galaxy SF distribution each pixel is found (note a very similar technique 
was independently developed by \citealt{fru06}).
Images are first trimmed to remove non-galaxy regions of the
field of view, and are then binned 3$\times$3 (in order to reduce the affects of
inaccurate SN coordinates and/or image alignment)\footnote{As the near-UV
images have an initial pixel scale of 1\as\ per pixel, these data were not
binned.}. Pixels within each image are then ordered in terms of
increasing count. From this ordered list we then form the cumulative
distribution. Each pixel is then
assigned an `NCR' value by dividing the cumulative pixel count by the total of
the cumulative distribution. All pixels forming part of the negative cumulative distribution are set
to an NCR value of zero (either sky or zero flux pixels).
Hence, all pixels have a value between 0 and 1,
where a value of 0 indicates that the pixel is consistent with zero flux or
sky values, whereas a pixel value of 1 means that the pixel has the
highest flux count within the image. To build up NCR statistics for
SN populations one simply takes the pixel where each SN falls and calculates
its corresponding NCR value.\\
\indent In the case of the current sample we apply this technique to each SN with respect to its
\ha\ image, and in addition with respect to its $GALEX$ near-UV image, broad-band
optical $B$ and $R$ images, and near-IR $J$- and $K$-band images. The
resulting NCR values are listed in Table A1. 
Any population which directly follows
the spatial distribution of light traced by each wave-band will show a
flat distribution of NCR values with a mean value of 0.5.
Whether a distribution follows such a one--to--one relation or not, then
provides constraints on progenitor properties. 
In the cumulative plots below if a distribution
accurately traces the light detected through a given filter, then one expects
the distribution to follow the diagonal one--to--one line shown on each plot.\\

\subsubsection{Spatial resolution}
During the above description of our data sets, we have detailed the pixel
scale of each distinct set of observations. During our processing of the host
galaxy images, most data are binned 3$\times$3 which means all images (UV through to near-IR) 
have NCR values which probe counts at a detector resolution of $\sim$1\as.
However, in some cases the seeing during observations was considerably worse than this. 
Of all observations the image quality in the $B$-band was the worst, hence we use these
data to present a limiting case of the spatial resolutions probed within
galaxies through our NCR statistic. The median seeing as measured on $B$-band
images was 1.6\as. The median recession velocity of our galaxy sample is 2647 \kms,
which equates to a distance 36.3 Mpc. These two values give a median spatial resolution
probed within galaxies of $\sim$280 pc. Hence, we resolve SN~Ia host galaxies
into a significant number of elements through our pixel statistics.

\subsection{Radial analysis}
While pixel statistics can give one information on the stellar population at
exact explosion sites, SNe~Ia have 
significant delay times and hence are likely to explode at considerable distances from
their birth sites. One method to further explore the environments of SNe is to 
investigate where they
explode with respect to the radial distribution of different stellar populations.
Within star-forming galaxies different age and metallicity stellar
populations are found at different characteristic galactocentric radial
positions. Therefore one can investigate where SNe are found within these more
generalised galaxy trends, and further infer progenitor properties.\\
\indent In \cite{jam06} we introduced a statistical radial analysis of host
galaxies, which compares the radial distribution of SNe to those of
stellar populations as traced by the continuum $R$-band light, and the \ha\ emission
respectively. To obtain these `\textit{Fr}' fractional flux values one
proceeds in the following fashion.
For each host galaxy elliptical apertures of increasing size are produced,
centred on the galaxy central coordinates, and calculated using the position
angle and ratio of major to minor axis of the galaxy (taken from NED). One
then finds the aperture which just includes the SN position. The flux (both of the
narrow and broad-band emission) within that aperture is then calculated, and
normalised to the total galaxy emission by dividing by the flux within an aperture
where the cumulative flux of the galaxy has become constant as one goes to
larger galactocentric radii. Hence, each SN has a radial value corresponding
to the continuum $R$-band light (\textit{Fr}$_{\textit{R}}$), and one
corresponding to \ha\ emission (\textit{Fr}$_{\textit{\ha}}$) between 0 and 1. A SN
having a \textit{Fr} value 0 would indicate that the SN exploded at the
central peak \ha\ or $R$-band pixel of the galaxy, while a value of 1 means that the
SN falls on an outer region where the galaxy flux is negligible above
the sky. \textit{Fr}$_{\textit{R}}$ and \textit{Fr}$_{\textit{\ha}}$ SN~Ia
distributions are hence built. If a SN population directly follows the
radial distribution of the broad- or narrow-band light, then the distribution
of \textit{Fr} values is expected to the flat. Therefore, we can investigate how
the different samples are associated with the radial positions of different
stellar populations. These techniques have been applied to CC SN samples 
in \cite{and09,hab10,hab12} and \cite{hab14}.\\

\section{Results}

\subsection{Pixel statistics}
\subsubsection{\ha}
We present the overall SNe~Ia \ha\ NCR pixel distribution with respect to
other main SN types in Figure 1 (note, an in-depth analysis of the CC SN sample was
presented in \citealt{and12}). It is immediately apparent
that SNe~Ia show the lowest degree of association with the on-going SF (as seen previously
by \citealt{bar94}, \citealt{jam06} and most recently \citealt{gal14}). 
This is to be expected
if, as generally assumed, SNe~Ia arise from some system containing a WD, i.e. a relatively old progenitor population.
The mean \ha\ NCR value (98 events\footnote{Note that not 
all SNe in the sample have both NCR and \textit{Fr} values, hence this value is lower than
the 102 stated for the full sample of events earlier.}) is 0.157, with a
standard deviation of 0.253, and a standard error on the mean of 0.026. This
compares to mean values of 0.254 (162.5 SNe), 0.303 (40.5) and 0.469 (52), for the
SN~II, SN~Ib and SN~Ic distributions respectively. 
Nearly 60\%\ of SNe~Ia within star-forming galaxies fall on zero \ha\ flux
(down to the limits of our observations, see \citealt{and12}).
Using a Kolmogorov-Smirnov test (KS-test) we find that there is less than a 0.1\%\ probability that
the SN~Ia and SN~II distributions are drawn from the same parent
population. The SN~Ia distribution does not
follow the on-going SF within galaxies. Given
that \ha\ emission is generally thought to be produced from the ionising flux
from stars less than 10 Myr old \citep{ken98}, this result is consistent with 
even the youngest predicted timescales for SN~Ia progenitors (see e.g. \citealt{aub08}). 

\begin{figure}
\includegraphics[width=8.5cm]{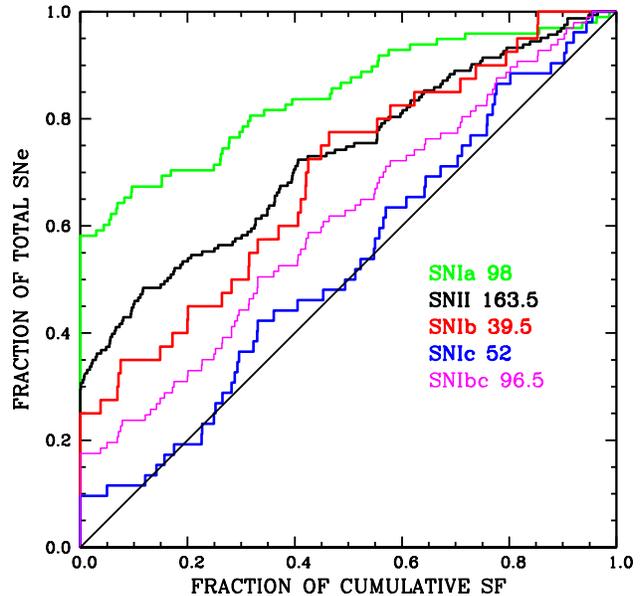}
\caption{
Cumulative \ha\ pixel statistics distributions of the main SN types. As distributions move 
away to the upper left from the black diagonal (a
hypothetical distribution, infinite in size, that accurately traces the
underlying on-going SF), they are showing a lower
association to the emission. (We note that this plot is the same as that presented 
in \citealt{and12}.)}
\end{figure}

\begin{figure}
\includegraphics[width=8.5cm]{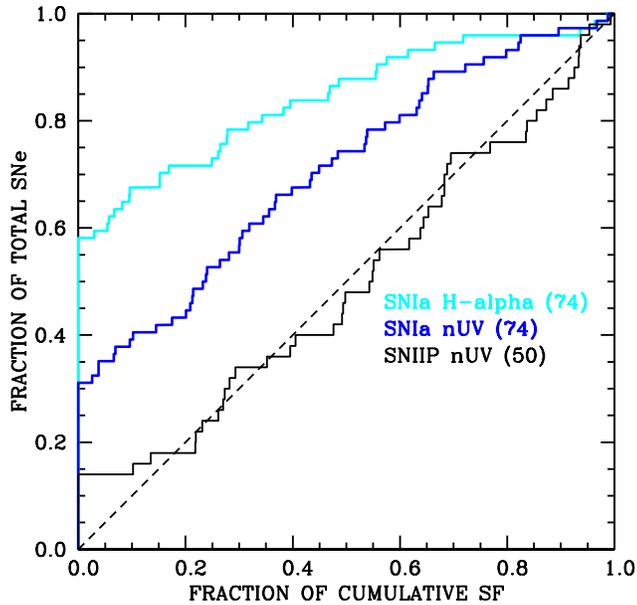}
\caption{Cumulative distribution NCR plot showing the SN~Ia population with respect to 
\ha\ and near-UV emission of their host
galaxies. For reference the SN~IIP distribution with respect to near-UV emission is
shown in solid black (taken from \citealt{and12}).}
\end{figure}

\subsubsection{Near-UV}
In Fig.\ 2 the cumulative NCR distributions are shown for 74 SNe~Ia, with respect to both
\ha, and \textit{GALEX} near-UV emission. The SN~Ia distribution
shows a higher degree of association to the UV emission than that of
\ha. However, it still appears that SNe~Ia do not accurately trace
recent SF. The mean near-UV NCR value is 0.292 with a standard deviation of 0.292 and a standard error on
the mean of 0.034. This compares to a mean \ha\ NCR value, for the same 74
events of 0.175 (standard deviation of 0.224, standard error on the mean of 0.026). 
Using a KS-test we find that there is a
0.5\%\ 
chance that the SN~Ia distribution shows the same degree of
association to both the \ha\ and near-UV emission. We also find
that there is less than 0.5\%\ chance that the SN~Ia and SN~IIP distributions,
with respect to the near-UV emission, are drawn from the same parent
population. Finally, the SN~Ia near-UV distribution has a less than
0.1\%\ chance of being drawn from a flat distribution, i.e. one that
accurately traces the recent SF. 
This result suggests that at least the majority of SNe~Ia found within star
forming galaxies do not explode on the timescales of $<$100 Myr traced by
near-UV emission \citep{gog09}.

\subsubsection{Multi wave-band NCR statistics}
In Fig.\ 3 we present the NCR distributions of SNe~Ia for multiple wavelength 
observations. The mean NCR values together with KS-test results of each distribution
between that wave-band and a flat distribution are presented in Table 2. The SN~Ia population 
best traces the $B$-band host galaxy light, followed
by the $R$-band light distribution. The SN~Ia population does
not follow the host galaxy light distribution in \ha, near-UV, nor the $J$- and $K$-bands. We
speculate that this indicates that the SN~Ia progenitor population in spiral galaxies does not
arise from either the very young populations of less than a few 100 Myrs (that traced by
\ha\ and near-UV emission), but neither is it dominated by long lived population of
several Gyrs (i.e. that traced by the $J$- and $K$-band light). The fact that the SN~Ia population
most closely follows the $B$-band light indicates that the progenitor population in
late-type galaxies is dominated by relatively young systems.\\
\indent It is also interesting to note how the different wave-band distributions compare to each other
with respect to the values of individual SNe. Using the Pearson's test for correlation, we find that nearly
all wave-bands NCR values show significant correlation\footnote{Where we define significant correlation as
a Pearson's $r$-value higher than 0.5.}. 
This means that if a SN has an NCR value near the top (or bottom) of the distribution in e.g. 
the $R$-band,
then it is likely to have an NCR value near the top (or bottom) of the distribution in e.g. \ha, although the absolute
NCR values may be significantly offset.
The only wavebands that do not show significant correlation
are: $K$ and near-UV; $K$ and \ha; and perhaps surprisingly near-UV and \ha. While 
a full discussion of the significance of these trends is beyond the scope of this paper, one
may speculate that this indicates that while there are differences between the stellar populations
traced by these different observations, the populations cluster together, i.e. where one
finds significant flux of stellar light (e.g. $B$- and $R$-band) one also finds
peaks of SF (\ha\ and near-UV). 
An analysis comparing pixel statistics between different wave-bands, and including 
all host galaxy pixels (i.e. not just those where SNe have exploded), may give
interesting constraints galaxy SFHs. Such analysis, independent
of SN studies, will be the focus of future work. 

\begin{table} \centering
\begin{tabular}[t]{cccc}
\hline
\hline
Imaging band & Number of SNe & Mean NCR & KS-test from\\
			 &               &          & flat distribution\\
\hline
\ha\ 	& 98	& 0.157 	& $<$0.1\% \\
near-UV & 74	& 0.292   & $<$0.1\% \\
$B$		& 43	& 0.498 	& $>$10\%  \\
$R$		& 80	& 0.415	  & $\sim$5\%\\
$J$		& 45 	& 0.345 	& $\sim$0.5\%\\
$K$		& 45    & 0.293 	& $<$0.1\% \\
\hline
\hline
\end{tabular}
\caption{NCR statistics for the 6 distinct wave-band analyses. In the first column the 
wave-band is indicated, followed by the number of SNe and host galaxies analysed in column 2. 
In column 3 the mean NCR value for each distribution is presented. Finally in column
4 the chance possibility of each distribution following a flat distribution is listed, as
calculated using the KS-test. (Note, if we only analyse those SNe common to all wave-bands, the results
are completely consistent with the above.)}
\end{table}

\begin{figure*}
\includegraphics[width=16cm]{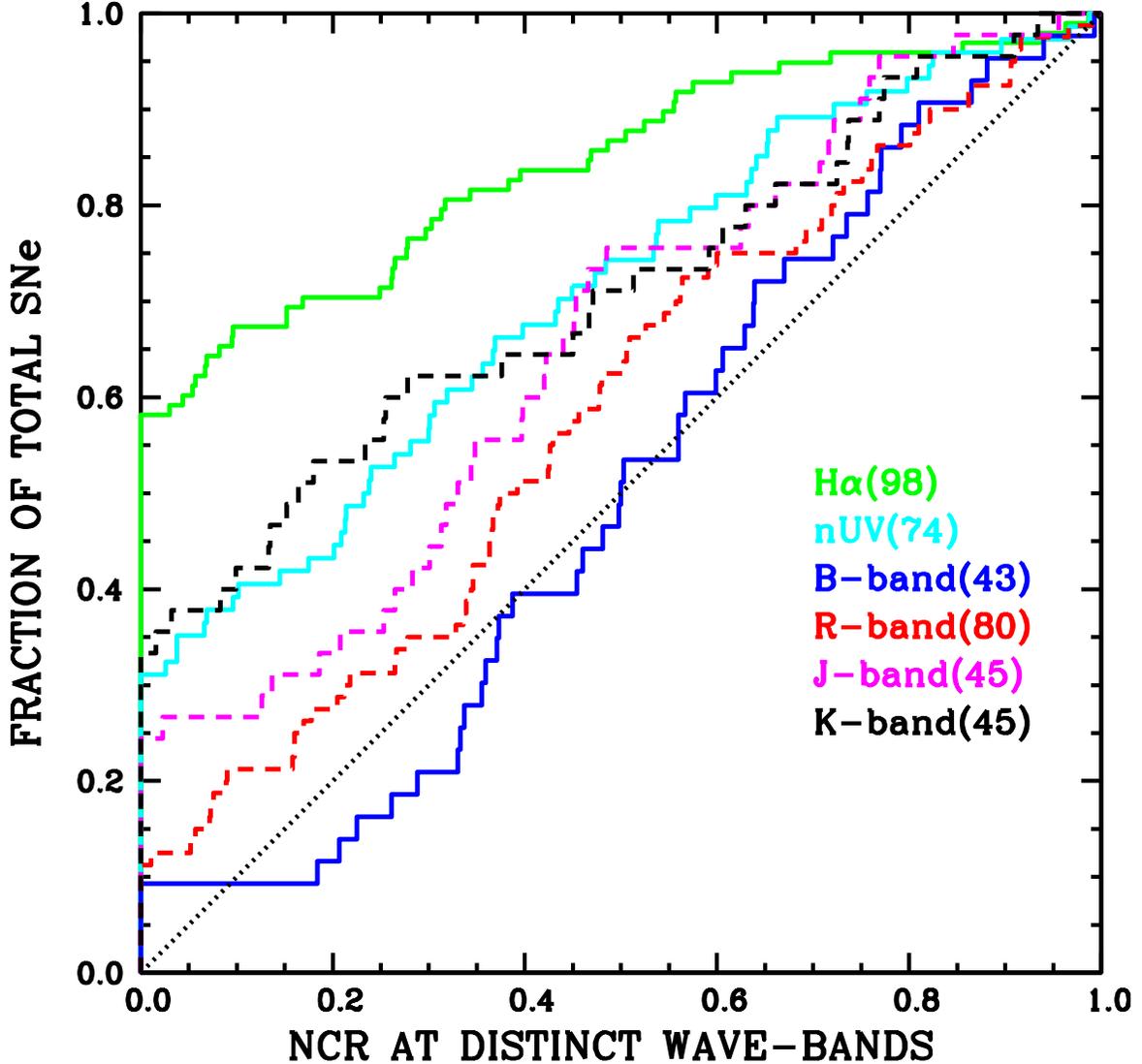}
\caption{NCR distributions for SNe~Ia with respect to 6 different wave-band host galaxy images (in
star-forming galaxies).}
\end{figure*}

\subsubsection{\ha\ pixel statistics with respect to event properties}
SNe~Ia light-curve stretch values were estimated for 56 SNe within the
sample, as outlined in \S\ 2.1. 
We split this
sample by the median stretch of 0.978 and
resulting \ha\ NCR cumulative distributions are shown in Fig. 4. 
No statistical difference between the two distributions is found.
We were able to estimate \textit{(B-V)}$_{max}$ values for 54 SNe and the sample is 
split by the median \textit{(B-V)}$_{max}$ value of 0.111.
The results from this analysis are presented in Fig. 5. `Redder' 
events (those with \textit{(B-V)}$_{max}$ $>$ 0.111) 
show a higher degree of association with \ha\ than
`bluer' SNe (similar to the results found by \citealt{rig13}). 
The mean \ha\ NCR for the `red' events is 0.209 (0.266, 0.052),
while for the `blue' events the mean is 0.070 (0.140, 0.027). 
Using a KS-test, this difference is significant at the $\sim$10\%\ level (in addition
the mean values are different at the 3 sigma level). An underlying
difference in the sense that `redder' SNe~Ia are more likely to be
found nearer/within bright \hii\ regions could be explained by the
fact that within
bright \hii\ regions SNe are likely to suffer from higher degrees of line of sight extinction. 
It is also possible that progenitor properties could play some role.\\

\begin{figure}
\includegraphics[width=8.5cm]{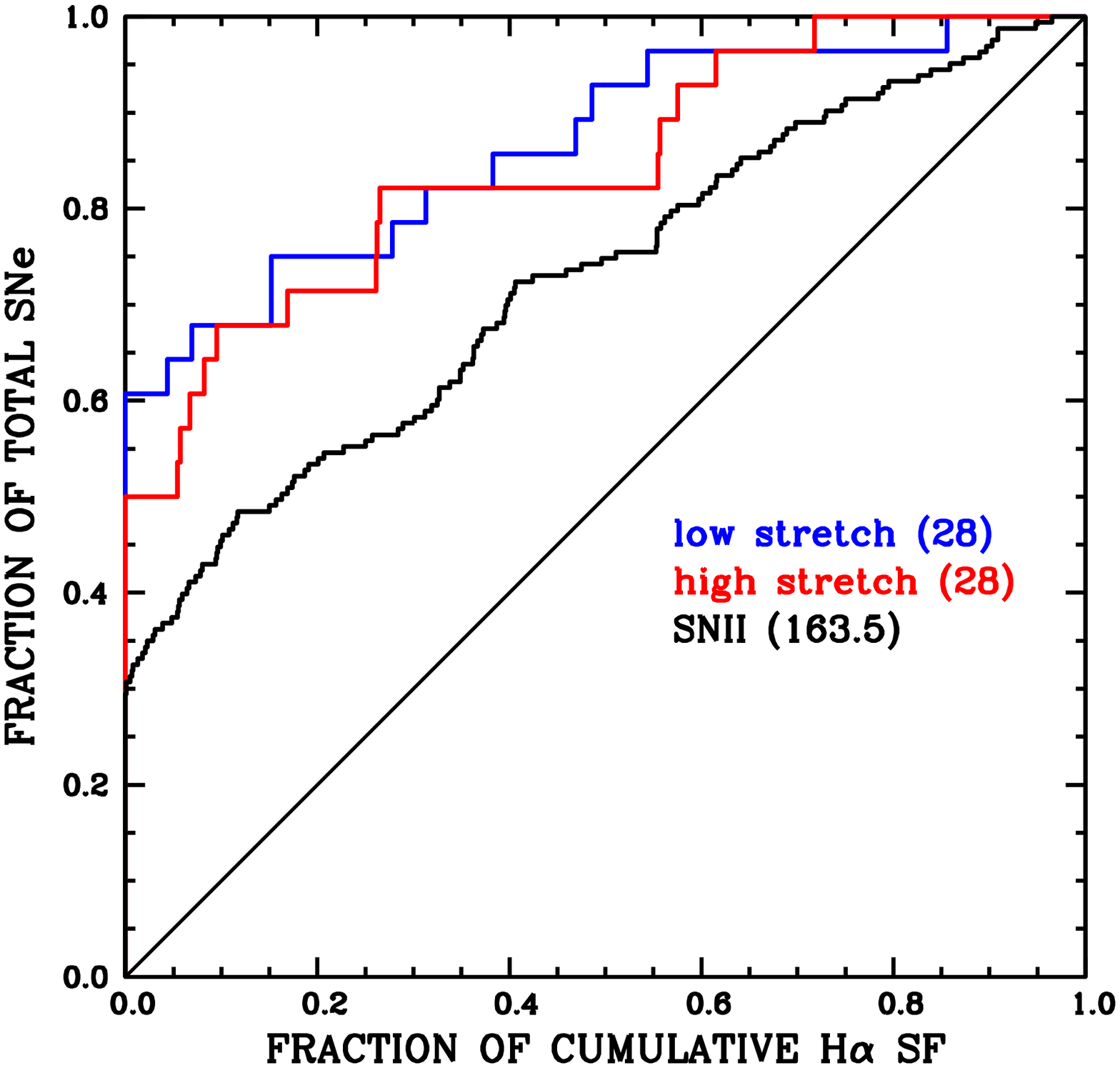}
\caption{\ha\ cumulative pixel statistics plot with SNe~Ia (in
star-forming galaxies) split into those above 
and below the median stretch value of
0.978. The SN~II distribution is shown for reference}
\end{figure}

\begin{figure}
\includegraphics[width=8.5cm]{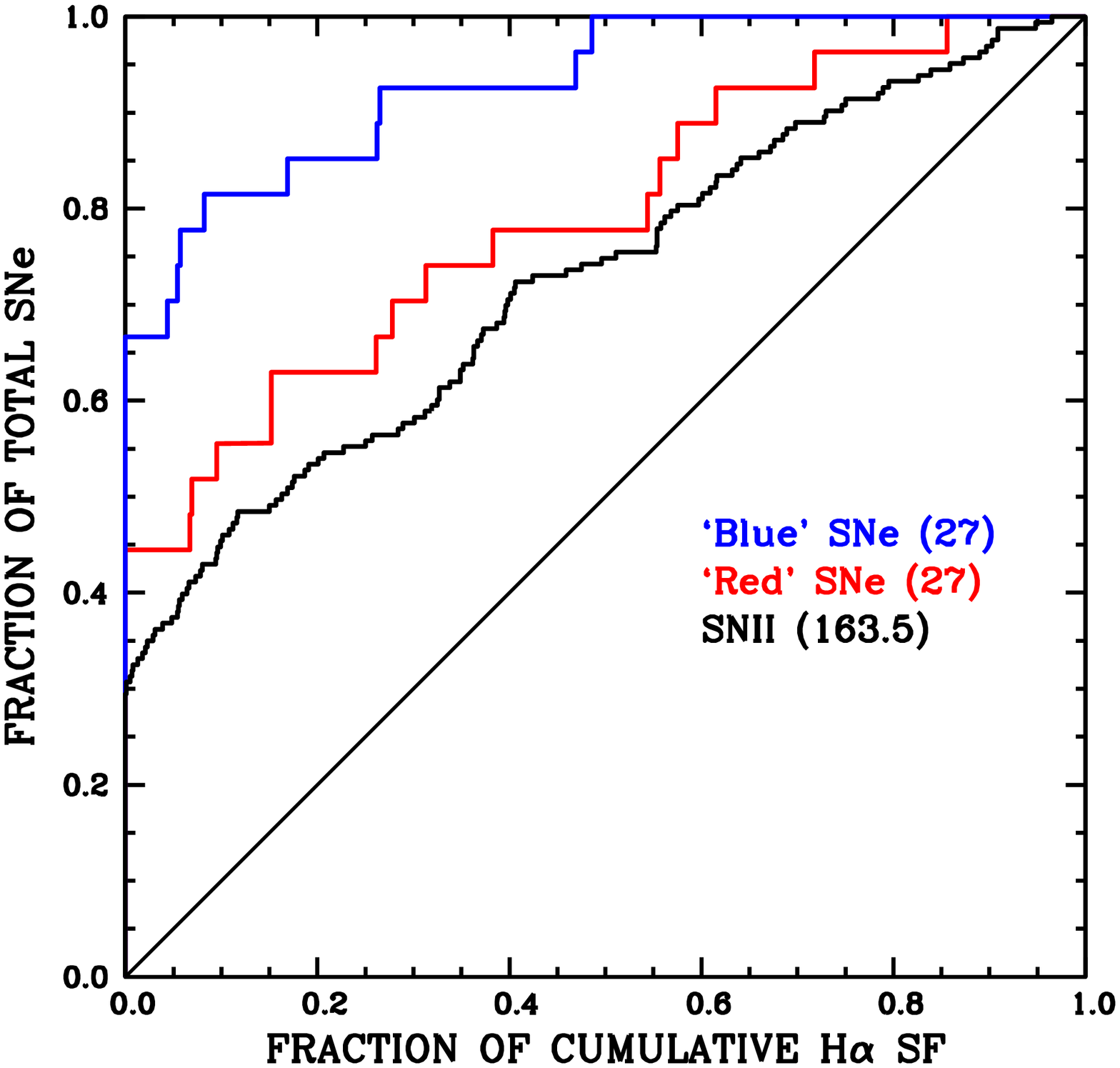}
\caption{\ha\ cumulative pixel statistics plot with SNe~Ia (in
star-forming galaxies) split into those above and
below the median \textit{(B-V)}$_{max}$ of 0.111. For reference the SN~II distribution is also shown.}
\end{figure}

\subsubsection{Near UV pixel statistics with respect to event properties}
Now we repeat the analysis presented in the previous section but with
respect to near-UV in place of \ha\ emission. There are some host 
galaxies within our \ha\ sample that do not have
corresponding near-UV images available. Therefore for
this sub-sample with available images we re-calculate median 
stretch and \textit{(B-V)}$_{max}$ values, finding
0.984 and 0.105 respectively. 
The mean near-UV NCR values for the stretch
distributions are 0.253 (0.267, 0.057) and 0.266 (0.307, 0.067), for the
high and low stretch SNe respectively, i.e. we find no statistical difference between 
the association of low and high stretch to recent SF, as shown in Fig.\ 6.
This suggests that even the most `prompt' SNe~Ia
found within star-forming galaxies --which are likely to
be related to the higher stretch population-- do not have delay times less than a few 100 Myrs.
The \textit{(B-V)}$_{max}$ distributions, when split by the median value of
0.105 are shown in Fig.\ 7. It appears that `redder' events
show a higher degree of association to the near-UV emission than
the `bluer' ones. Again this trend is probably most easily explained by these SNe
suffering from higher extinction from line of sight material, but
we stress the possibility of progenitor properties playing some role. Below we further investigate the environments
of the `reddest' events in our sample.
We note that if we repeat the above analysis with respect to the other wavebands analysed above we do
not observe any significant differences between the stretch or colour samples.\\

\begin{figure}
\includegraphics[width=8.5cm]{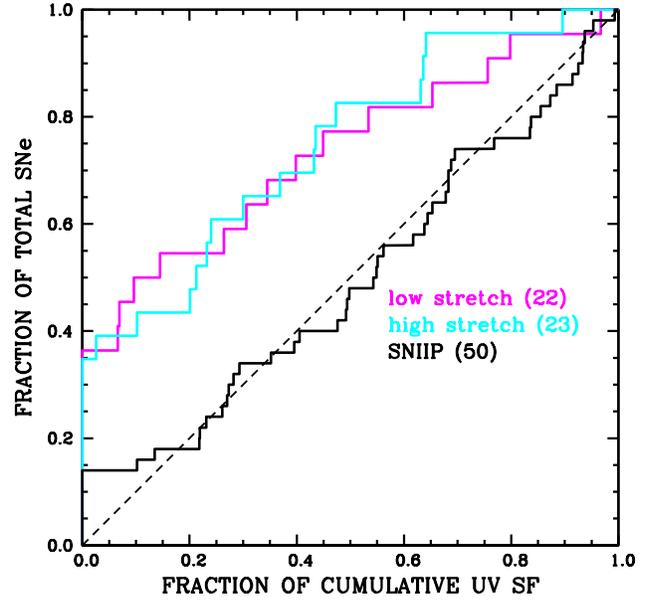}
\caption{\textit{GALEX} near-UV cumulative pixel statistics plot with SNe~Ia 
(in
star-forming galaxies) split into those above 
and below the media stretch value of the sample of
0.984. 
The SN~IIP distribution (with respect to near-UV emission) is shown for
reference}
\end{figure}

\begin{figure}
\includegraphics[width=8.5cm]{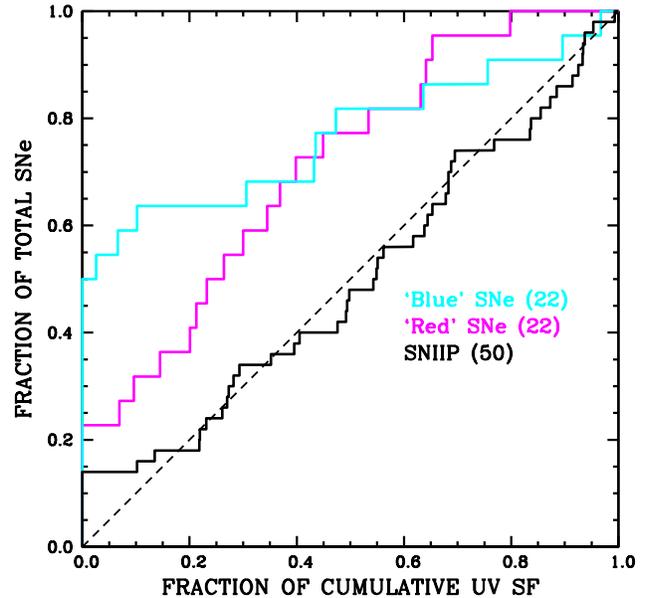}
\caption{\textit{GALEX} near-UV cumulative pixel statistics plot with SNe~Ia 
(in
star-forming galaxies) split into those above and
below the median \textit{(B-V)}$_{max}$ of 0.105. The SN~IIP distribution is shown for
reference}
\end{figure}

\subsection{Radial analysis}
Here we present results from the fractional radial analysis,
the methods for which were outlined in \S\ 3. We ask two
main questions: 1) does the radial distribution of SNe~Ia within host galaxies
accurately trace the $R$-band stellar continuum 
population, or more accurately the \ha\ emission, a tracer of the on-going SF
within host galaxies? 2) If we split the SN~Ia distribution by light-curve
parameters, do specific types of SNe~Ia occur at different characteristic radial
positions within galaxies? We start by analysing the overall radial distributions\footnote{We do not present 
\textit{Fr} distributions with respect to the other wave-bands analysed through our
NCR method above. While pixels statistics give specific information
at the exact location of each SNe, the \textit{Fr} analysis is a much more crude environment
property estimator, where differences between similar wave-length observations are not
significant. In order
to not overload the reader with additional analysis and figures we choose present only
the \ha\ and $R$-band analysis in this section.}. 

\subsubsection{Overall SN~Ia radial distributions}
In Fig. 8 a histogram of the radial distribution of SN~Ia with respect to
the stellar continuum of their hosts is presented. The
mean \textit{Fr}$_{\textit{R}}$ for the overall SN~Ia distribution (99 SNe) is
0.554 (standard deviation of 0.261, standard error on the mean of
0.026). Hence, the distribution appears to be slightly
biased towards larger radial distances when compared to the $R$-band
light. There appears to be a central deficit of SNe~Ia, and indeed this deficit is seen when we
apply a KS-test between the SN~Ia \textit{Fr}$_{\textit{R}}$ and a flat
distribution. We find a $\sim$7\%\ chance that the SNe~Ia 
accurately traces the radial $R$-band distribution of their hosts. The
easiest way to interpret this result would be that it is solely a selection effect
where SNe~Ia go undetected in the central parts of galaxies due to higher levels of extinction and higher surface
brightness. However, in Fig. 9 we show the SN~Ia \textit{Fr}$_{\textit{R}}$, plot in a
cumulative distribution, compared to that of 185 SNe~II and 97 SNe~Ibc (these
results have been presented in \citealt{and09,hab10,hab12}). 
This shows that while the SNe~II also have a central deficit, SNe~Ibc do not show any such deficit. 
Given that SNe~Ia are
generally more luminous than SNe~Ibc (meaning that any selection effect against finding SNe
in the central parts of galaxies should be worse for the latter), this suggests that
the lack of SNe~Ia found within the central parts of galaxies is a real, 
intrinsic observation.\\
\indent In Fig. 10 we show a histogram of the radial distribution of SNe~Ia compared to
the distribution of on-going SF of their hosts. The
mean \textit{Fr}$_{\textit{\ha}}$ is
0.560 (0.313, 0.031). As for the distribution with respect to the $R$-band
light, the SNe~Ia appear to explode at slightly larger galactocentric distances
with respect to \ha\ emission. However, there is no large deficit in the
central parts of the emission. This
is also seen using a KS-test where the SN~Ia distribution is consistent with
being drawn from a flat distribution, i.e. one that accurately traces the
radial distribution of \ha\ emission.
In Fig. 11 the SN~Ia radial \ha\ cumulative distribution
is presented, compared to that of 185 SNe~II and 97 SNe~Ibc. SNe~Ia
appear to become more numerous per unit SF out to larger galactocentric
radii than the CC SNe within our previously published samples.\\

\begin{figure}
\includegraphics[width=8.5cm]{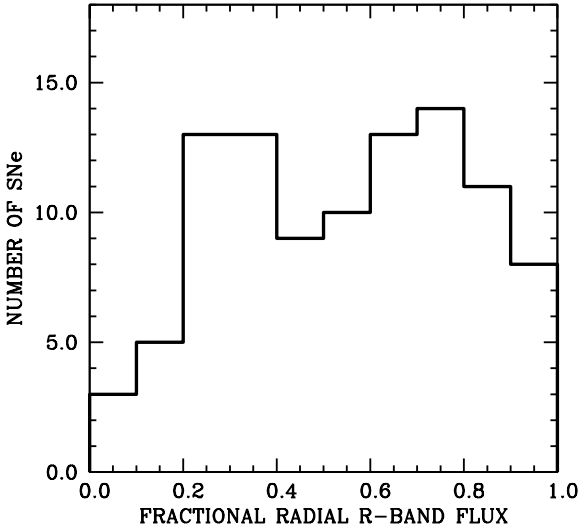}
\caption{Histogram of fractional radial distribution of the overall SN~Ia
population (99 SNe in
star-forming galaxies) with respect to the $R$-band light distribution of their
hosts.}
\end{figure}

\begin{figure}
\includegraphics[width=8.5cm]{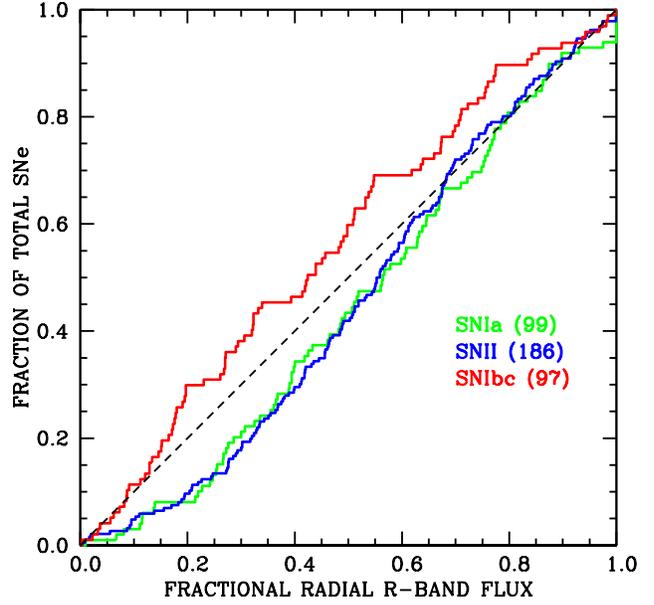}
\caption{Cumulative distribution of the \textit{Fr}$_{\textit{\R}}$ values of SNe~Ia (in
star-forming galaxies) and the CC SN types, SN~II and SN~Ibc, 
with respect to the $R$-band light distribution within their host
galaxies.}
\end{figure}

\begin{figure}
\includegraphics[width=8.5cm]{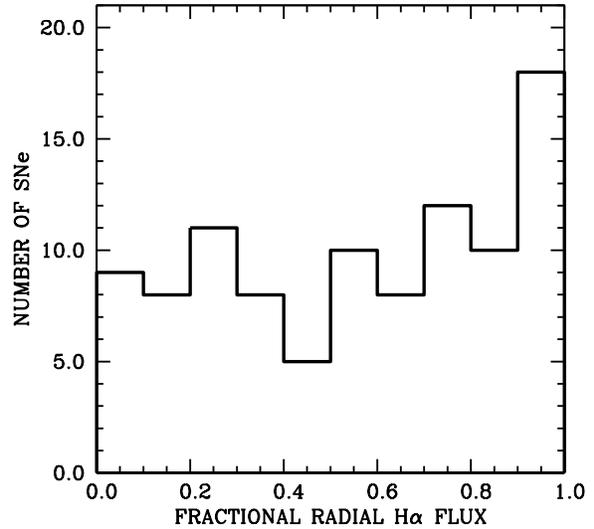}
\caption{Histogram of fractional radial distribution of the overall SN~Ia
population (99 SNe in
star-forming galaxies) with respect to the \ha\ line emission distribution of their
host galaxies.}
\end{figure}

\begin{figure}
\includegraphics[width=8.5cm]{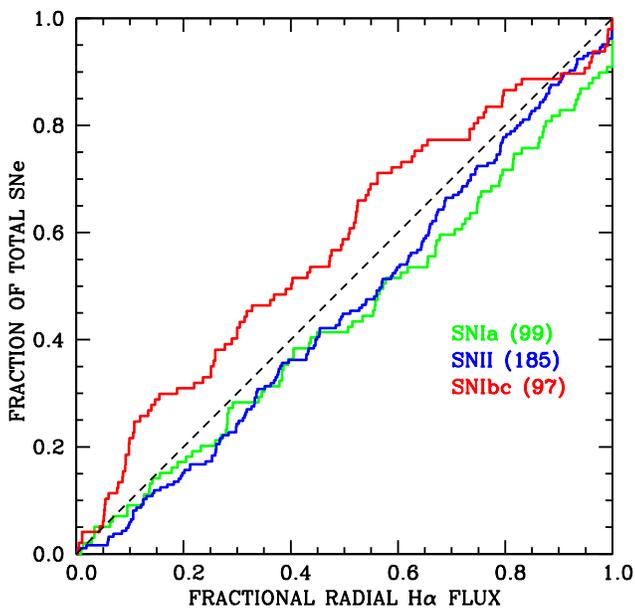}
\caption{Cumulative distribution of the radial distribution of SNe~Ia (in
star-forming galaxies) and the CC SN types, SN~II and SN~Ibc,
with respect to the \ha\ emission distribution within their host
galaxies.}
\end{figure}

\subsubsection{Radial distributions of SNe~Ia when split by light curve parameters}
As was done for the pixel statistics analysis above, we now analyse whether
there are differences in the radial distributions of SNe~Ia when the sample is
split by light-curve parameters. The sample is split by a median 
stretch of 0.984, and by the median 
\textit{(B-V)}$_{max}$ of 0.106. In Figs. 12 and 13 histograms of the radial
distributions of SNe~Ia with respect to the $R$-band and \ha\ light are shown,
with the sample split by stretch. In 
Figs. 14 and 15 histograms of the radial
distributions are presented
with the sample split by \textit{(B-V)}$_{max}$. The mean values
(together with standard deviations and standard errors on the mean), plus
KS-test percentage values, and sigma differences between means are listed
in Table 3.\\
\indent The most significant difference between these various populations
is between the \textit{Fr}$_{\textit{R}}$ distributions when split by SN colour.
A KS-test shows the distributions to be different with a 0.3\%\ chance probability
of being drawn 
from the same underlying population. This difference is in the sense that `redder' events are found more
centrally within host galaxies, as also observed by \cite{gal12}.

\begin{figure}
\includegraphics[width=8.5cm]{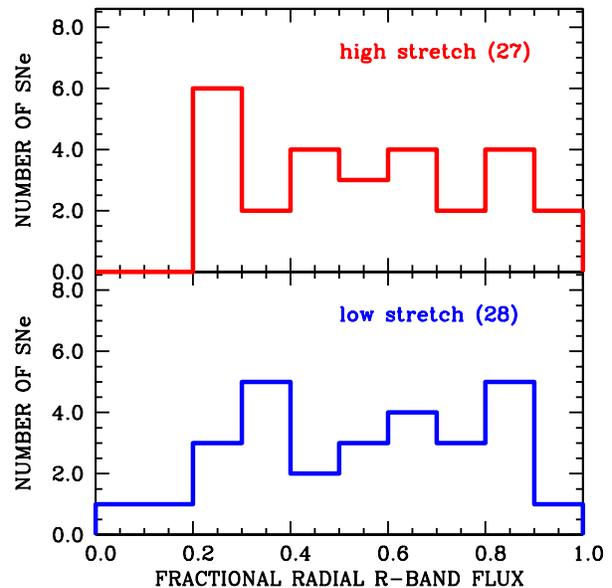}
\caption{Histogram of fractional radial distribution of the SN~Ia
population (in
star-forming galaxies) with respect to the $R$-band light distribution of their
host galaxies, when split by the median stretch
value of 0.984.}
\end{figure}

\begin{figure}
\includegraphics[width=8.5cm]{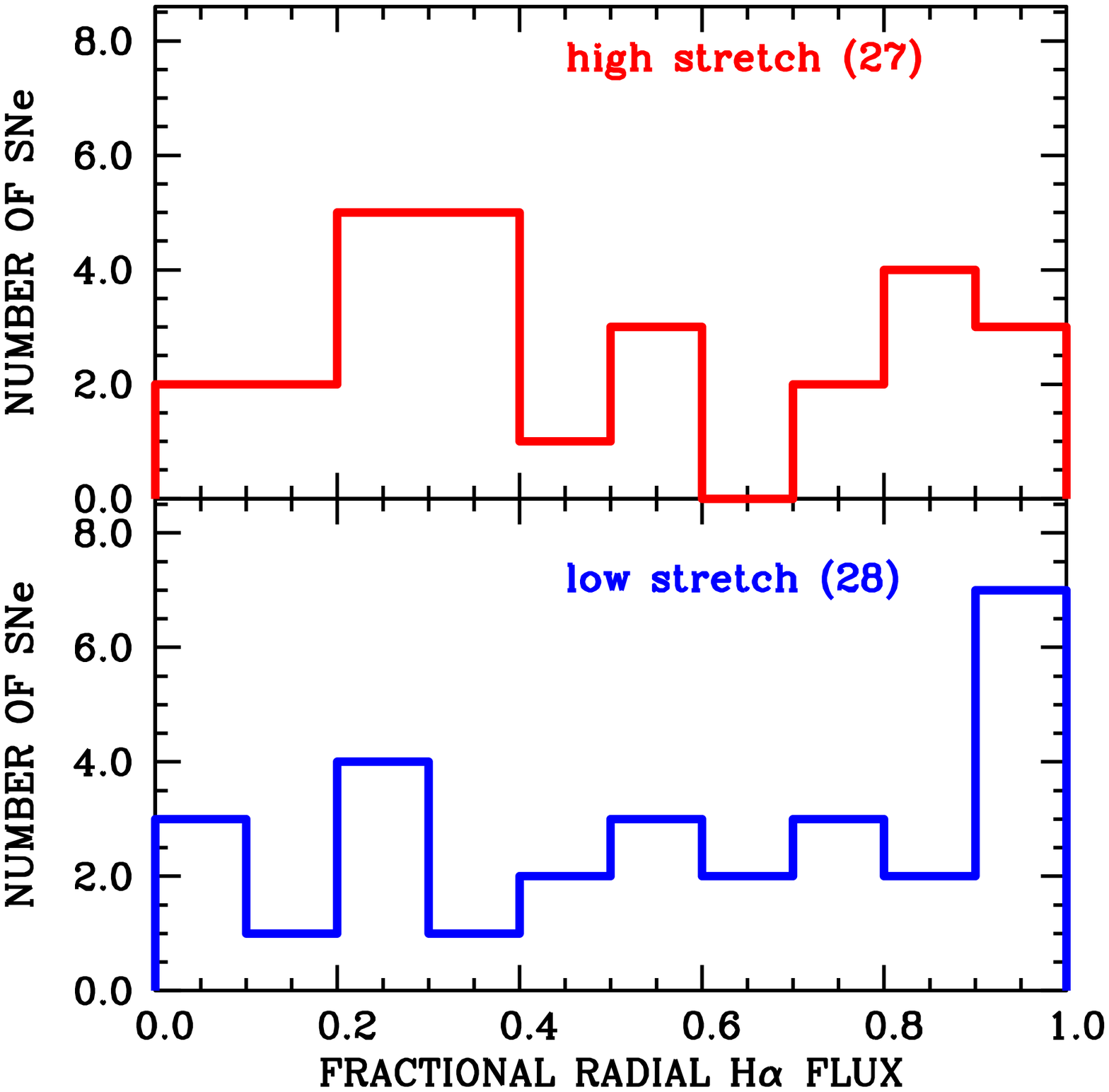}
\caption{Histogram of fractional radial distribution of the SN~Ia
population (in
star-forming galaxies) with respect to the \ha\ emission distribution of their
host galaxies, when split by the median stretch
value of 0.984.}
\end{figure}

\begin{table*}
\begin{tabular}[t]{ccccc}
\hline
\hline
Distribution & N. of SNe & Mean \textit{Fr} (stdev, sterr) & KS-test &sigma difference\\
\hline
\textit{Fr}$_{\textit{R}}$ high stretch & 27 & 0.562 (0.241, 0.047) & $>$10\%\ & 0.08\\
\textit{Fr}$_{\textit{R}}$ low stretch & 28 & 0.557 (0.248, 0.048) & & \\ 
\hline
\textit{Fr}$_{\textit{\ha}}$ high stretch& 27 & 0.498 (0.302, 0.059) & $>$10\%\ & 0.97\\ 
\textit{Fr}$_{\textit{\ha}}$ low stretch  &28&0.581 (0.323, 0.061) & &\\  
\hline
\textit{Fr}$_{\textit{R}}$ red \textit{(B-V)}$_{max}$ & 26 & 0.500 (0.245, 0.041) & 0.3\%\ & 1.90\\
\textit{Fr}$_{\textit{R}}$ blue \textit{(B-V)}$_{max}$ &26 & 0.619 (0.234, 0.047) & & \\
\hline
\textit{Fr}$_{\textit{\ha}}$ red \textit{(B-V)}$_{max}$ &26 & 0.460 (0.311, 0.063) & $>$10\%\ & 1.90\\
\textit{Fr}$_{\textit{\ha}}$ blue \textit{(B-V)}$_{max}$& 26& 0.625 (0.308, 0.062) & &\\

\hline
\end{tabular}
\caption{Mean \textit{Fr} values for each of the 8
distributions, with respect to both the $R$-band and \ha\ light. In column 1
the distribution name is given, followed by the number of events within that
distribution in column 2. Then the mean \textit{Fr} values are given
together with their associated standard deviations and standard errors on the
mean. Finally for each set of distributions (i.e. split by either light-curve
stretch or colour) we list the KS-test probabilities that the two
distributions are drawn from the same parent population, followed by the
statistical difference (in terms of sigma) between the mean values of each
population.}
\end{table*}

\begin{figure}
\includegraphics[width=8.5cm]{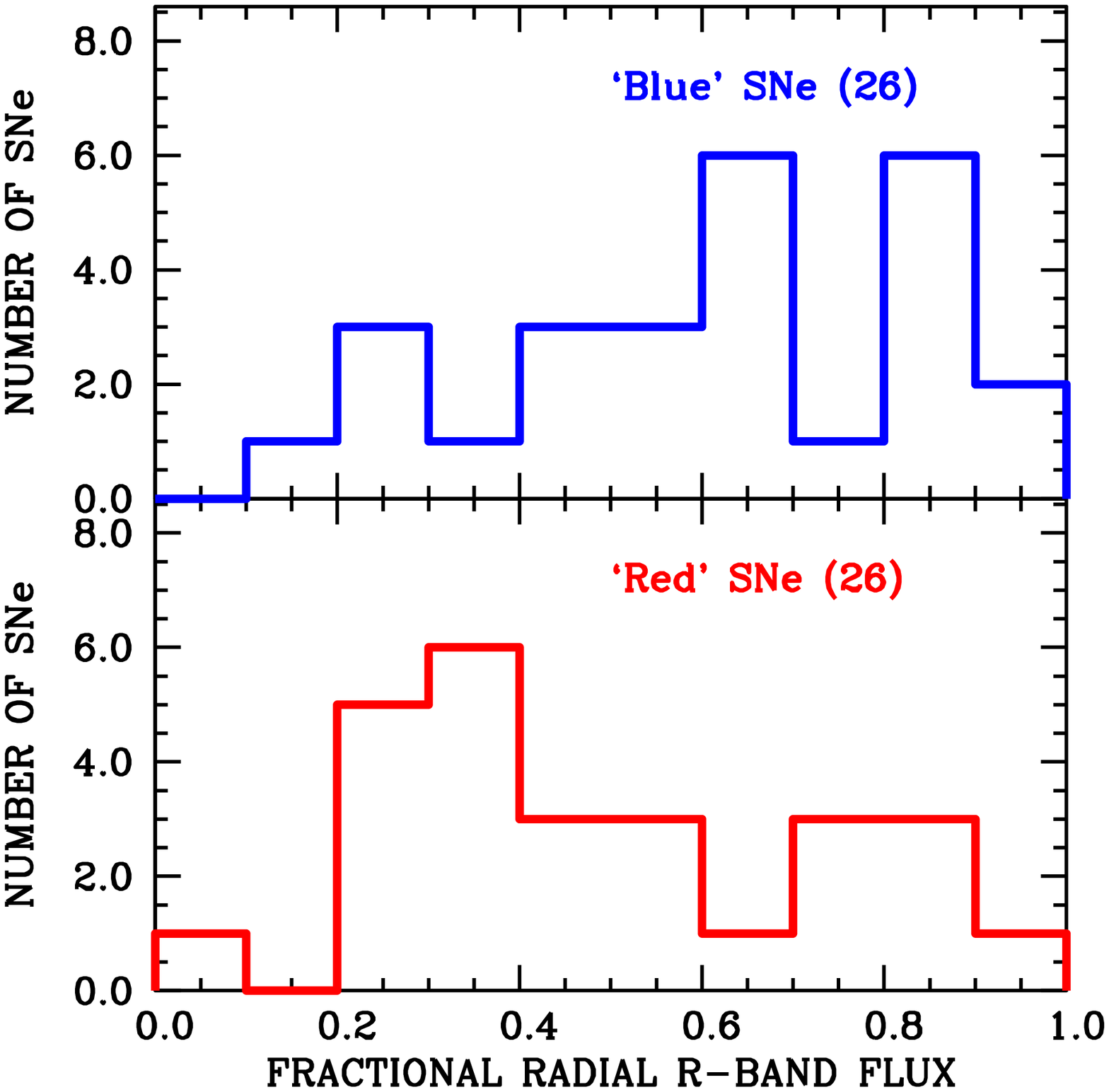}
\caption{Histogram of fractional radial distribution of the SN~Ia
population (in
star-forming galaxies) with respect to the $R$-band light distribution of their
host galaxies, when split by the median \textit{(B-V)}$_{max}$ of 0.106. }
\end{figure}

\begin{figure}
\includegraphics[width=8.5cm]{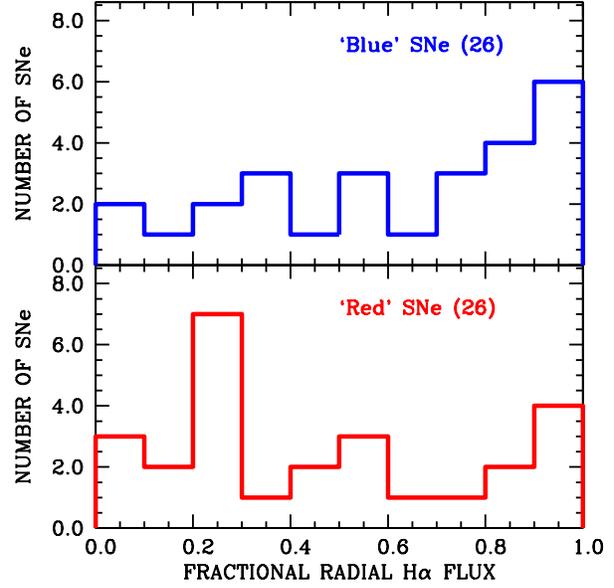}
\caption{Histogram of fractional radial distribution of the SN~Ia
population (in
star-forming galaxies) with respect to the \ha\ emission distribution of their
host galaxies, when split by the median \textit{(B-V)}$_{max}$ of 0.106. }
\end{figure}

\subsection{The environments of the `reddest' SNe~Ia}
It has been shown that `redder' SNe are generally found to both fall on
regions of more intense SF and in more central parts of galaxies than their `blue' counterparts.
If these results are due to line of sight ISM then one may ask if there are 
SNe which are significantly `red' but \textit{do not}
fall on star-forming regions or in the central parts of their galaxies?
We define `red' SNe as those with a \textit{(B-V)}$_{max}$
value higher than 0.5. We then compile the data for these SNe in Table 4 where we present 
SN colours together with their \ha\ and near-UV NCR values, and their \textit{Fr}$_{\textit{R}}$ and 
\textit{F}$_{\textit{\ha}}$ values.
We find that these `reddest' SNe indeed generally occur within bright \hii\ regions (as indicated by their NCR values)
and in more central regions. The mean \ha\ NCR value for these 10 SNe is 0.232, 
higher than the 0.157 for the full sample. In addition, almost 60\%\ of the full sample
fall on regions of zero \ha\ flux, while this falls to 40\%\ for these 10 SNe.
In terms of their radial positions, the mean \textit{Fr}$_{\textit{R}}$ for this subset of events is 0.408, compared to
0.554 for the full sample, and the difference is more extreme for \textit{Fr}$_{\textit{\ha}}$: 0.292 for the 
sub-sample, and 0.560 for the full sample. Looking closely at Table 4
there is only one SN where one may argue that the environmental properties cannot explain 
its `red' colour: SN~2005ke (the rest of the sample either fall on regions of significant SF
or in central regions).\footnote{Indeed, other than this SN \textit{all} of the events within this
sub-sample exploded within the central 50\%\ of their host galaxy light.}
We note that this is the `bluest' SN within this sub-sample.
Below we further discuss the implications of this result.

\begin{table} \centering
\begin{tabular}[t]{cccccc}
\hline
\hline
SN & \textit{(B-V)}$_{max}$ & NCR \ha  & NCR UV & \textit{Fr}$_{\textit{R}}$ & \textit{Fr}$_{\textit{\ha}}$ \\
\hline
1986G & 0.839	& 0.069	& $\cdots$		&$\cdots$			&	$\cdots$	   \\
1995E & 0.679	& 0.000 & 0.345	& 0.394 & 0.284\\
1996ai& 1.553   & 0.615 & 0.232 & 0.325 & 0.292\\
1999bh& 0.872   & 0.856 & $\cdots$	& 0.418& 0.232\\
1999cl& 1.059   & 0.152 & 0.264 & $\cdots$&$\cdots$\\
2003cg& 1.110   & 0.557 & 0.369 & 0.242 & 0.142\\
2005A & 0.986   & 0.000 & 0.398 & 0.255 & 0.136\\
2005ke& 0.653   & 0.000 & 0.069 & 0.861 & 0.815\\
2006X & 1.196   & 0.067 & 0.201 & $\cdots$&$\cdots$\\
2007N & 0.988   & 0.000 & 0.000 & 0.361 & 0.282\\
\hline
\hline
\end{tabular}
\caption{The environment statistics for the `reddest' SNe~Ia in the sample.
Here we show all SNe which have \textit{(B-V)}$_{max}$ higher than 0.5. In the first column the 
SN name is listed, followed by the \textit{(B-V)}$_{max}$ in column 2. In columns
3 and 4 we present the NCR values derived from \ha\ and near-UV imaging respectively.
Then the \textit{Fr}$_{\textit{R}}$ and 
\textit{Fr}$_{\textit{\ha}}$ values are listed in columns 5 and 6 respectively.}
\end{table}

\section{Discussion}
In previous sections statistical distributions of SN~Ia environments 
have been presented together with results characterising the properties of the stellar
light found in the vicinity of SNe~Ia within star-forming galaxies. Now we 
further discuss how these results can be understood in terms of progenitor
properties and SNe~Ia diversity. 

\subsection{Implications for SN~Ia progenitors} 
It has long been accepted that SNe~Ia found within star-forming galaxies
have shorter lifetimes than those found within elliptical galaxies, due to the 
longer lived population of the latter. Indeed, there is now
significant evidence in the literature that late-type galaxies host brighter SNe~Ia
\citep{ham00,sul06,kel10,sul10,lam10,gup11,dan11,joh13,hay12,chi13,pan14}, 
and these events are generally assumed
to form any `prompt' progenitor channel. 
In the current work we have only included star-forming host
galaxies within our sample. This is by design, as our main initial goal was to investigate 
the association of SNe of all types with host
galaxy SF. Hence, we remove a large fraction the SN
population which is distinct from that found in our sample. However, late-type 
galaxies have much more diverse stellar populations than those found in
ellipticals, and hence one may hope that environmental differences exist that can 
be used to constrain progenitor properties. These involve the ages traced by different
wave-bands analysed, together with metallicity gradients within spiral galaxies, and
in addition the wide range of line of sight extinction found within different regions of
star-forming galaxies.

\subsubsection{Constraints on SNe~Ia progenitor ages}
SN~Ia do not trace the \ha\ nor the near-UV emission, i.e.
the on-going or recent SF. However, they do seem to trace the $B$-band light extremely well (see Fig.\ 3). 
Going further redwards they do not trace the $J$- or $K$-band light.
Qualitatively, this suggests that the dominant progenitor population in late-type galaxies is neither
extremely prompt, nor significantly delayed. We speculate that the population is dominated 
by progenitors with ages of several 100s of Myrs.
As our technique is statistical in nature, we cannot
rule out extreme young or extreme old progenitors for any given SN.
However, our results suggest that the $B$-band
light distribution within star-forming galaxies is consistent with the
stellar population which traces the peak of the SN~Ia delay time distribution (DTD). 
Indeed, it has been argued by \cite{chi14} that this
where the great majority of SNe~Ia will arise from within these galaxies.
\ha\ and near-UV emission trace stellar populations with ages of less than around 10 and 100 Myrs respectively.
The fact that the SN~Ia population investigated here does not follow the
light of either tracer of SF
constrains the majority of these progenitors to have delay-times significantly longer than
these limits.
The observation that SNe~Ia in star-forming galaxies also does
not trace the near-IR light distribution --as traced by $J$- and $K$-band observations--
would also appear 
to constrain the majority of their progenitors to have significantly
shorter delay-times than traced by these near-IR observations.
While the SN population shows some degree of correlation with 
the $R$-band light, it best follows the $B$-band light. 
We use the output from
the population synthesis models of \cite{pie04} to extract the population age
where the peak flux coincides with the central wavelength of $B$-band filter
observations (note, other such models give very similar results). We find that 
a population age of $\sim$750 Myrs matches well the wavelength range probed by
$B$-band observations (which is demonstrated with a different set of models in figure 9
of \citealt{bru03}). Hence, this constrains the peak age of progenitors
within our star-forming galaxy sample to fall within a similar range.\\
\indent \cite{ras09}, using similar pixel techniques to those used
here (but confined to regions in the immediate vicinity of the SN), compared SN~Ia environments
to analytical galaxy models, concluding that even the `prompt' SNe~Ia 
progenitor channel exhibits a delay time of 200--500 Myrs, consistent with our more qualitative 
arguments above. An important point here is that \cite{ras09} also commented that SNe~Ia
and CC~SNe show a similar degree of association to the $g$-band light of their host galaxies. However,
while we make no comparison between the two SN types with respect to $g$-band light (or more appropriately,  
$B$-band as analysed here), in Fig.\ 2 it is clear that SNe~Ia do not follow the near-UV emission 
while SNe~II (and other CC types to a higher degree, see \citealt{hab14}) show almost a perfect
one-to-one relation to the recent SF. Hence, this shows that indeed one can separate the two SN types
using multi-wavelength pixel statistics analysed for complete galaxies.\\
\indent While the above argues against a progenitor population dominated by extremely young channels 
for the overall sample analysed, one may speculate that the `prompt' channel is only a fraction of our sample. 
In \S\ 4.1 we indeed separated the sample by `stretch',
where one would assume that the SNe with larger values (i.e. brighter events) are most
likely to form the `prompt' channel\footnote{We note that given our
sample lacks any elliptical galaxy hosts, it is essentially dominated by moderate--stretch
`prompt' SNe in the traditional classification.}. However, there is no difference in the association
of the two SN groups with either on-going or recent SF. 
Hence, even the majority of the most `prompt' SNe~Ia in star-forming galaxies
would appear to be constrained to have progenitors 
with delay times longer than at least 100 Myrs. We note, the
statistical nature of our analysis does not rule out any particular
SN having shorter lifetimes. However, we suggest that the
relative rate of any population (if it indeed exists) must be quite small.

\subsubsection{Radial distributions of SNe~Ia: progenitor age or metallicity constraints?}
In \S\ 4.2 it was shown that the central parts of galaxies are under-populated by SNe~Ia, 
and that it appears unlikely that this is a selection effect given that 
SNe~Ibc --which are intrinsically dimmer than SNe~Ia-- do not show any such deficit. 
This central deficit of SNe~Ia with respect to CC SNe was also previously
shown by \cite{wan97}.
In addition, as one goes out to larger galactocentric normalised distances there is a suggestion that
per unit SF more SNe~Ia are being produced (see Fig.\ 10). 
One can attempt to explain these results through both progenitor age and metallicity effects. In this
section we explore both these possibilities.\\
\indent A progenitor metallicity effect may manifest itself, when one considers
that metallicity gradients
are found within galaxies (central regions having higher abundances than outer regions,
see e.g. \citealt{hen99}). 
Thus the above results could be interpreted in that SNe~Ia prefer to explode within lower abundance regions of galaxies. 
Indeed there is some suggestion 
of such a trend in recent observational and theoretical works. \cite{pri08b}
concluded that there is no significant low-metallicity
threshold (below which SNe~Ia are not produced) by investigating global host metallicities. 
\cite{li11} showed
that the SN~Ia rate per unit mass has a strong relation to host galaxy mass, with 
more events being produced per unit mass in lower mass galaxies. \cite{qui12} also
found an excess of dwarf hosts (i.e. low metallicity) in a sample produced
by the non-targeted search of ROTSE-IIIb. These observational results were
then used by \cite{kis13}, who argued that lower metallicity progenitors will lead
to higher WD masses which then elevates the SN~Ia rate in low metallicity environments as a higher fraction of
WDs are able to explode as a SN~Ia. Indeed, a prediction
of \cite{kis13} was that the SN~Ia rate should be higher in the outer regions of galaxies, 
which is suggested by our work.
It is important to note at this point the earlier theoretical work which in fact predicted the
opposite to the above. \cite{kob98} proposed that SD progenitors should show a preference for
high metallicity because of a lower metallicity limit for the production of SNe~Ia. This is related
to the strong wind blown by the accreting WD, which was required for the progenitor 
to reach the Chandrasekhar mass. If metallicity is too low then the wind is too weak 
for the progenitor to evolve to explosion \citep{kob98}. Observations (including our own) appear
to rule out such a distinct metallicity effect. However, additional work is
needed to further investigate this issue. One possibility is to measure the metallicity
at the exact explosion site of SNe~Ia. If SNe~Ia indeed prefer higher/lower metallicity
then one would expect that their explosion sites would have higher/lower values than 
average regions throughout their galaxies. Such analyses are now possible with the advent of
wide field of view integral field spectrographs. It has also been shown that `redder' events are found more centrally within galaxies. 
While we discuss this below in terms of progenitor age, and then line of sight extinction effects, 
it is also
possible that this could be explained through a metallicity effect, which would imply
that `redder' SNe arise from higher metallicity progenitors.\\
\indent Progenitor age could also be the explanation for our radial distribution results. Within star-forming
spiral galaxies, in addition to metallicity gradients, one observes significant gradients
in population SFHs. This is in the sense that the outer regions have SFHs dominated by younger populations,
while more central regions have a much higher fraction of old stars.
It has been observed that the SN~Ia rate per unit mass is highly dependent
on the colour of parent populations (galaxies), with \cite{man05} showing that
the rate per unit mass is much higher in galaxies with bluer colours. Hence,
redder, older, more central stellar populations within galaxies have lower
SN~Ia rate per unit mass, providing an explanation for the lack of 
SNe in the central parts of the $R$-band light distribution of hosts.
Indeed, this argument can also be used to explain that the elevated rate of SN~Ia in
low mass galaxies is due to a population age effect (rather
than metallicity), as these galaxies are more actively star-forming, and hence
SN~Ia progenitor ages peak at the peak of the DTD, as outlined in \cite{chi14}. 
If this second effect of progenitor age is the dominant factor which
explains SN~Ia environment properties, then one may argue that metallicity 
plays only a minor role. Indeed, there are now several investigations
which argue that progenitor population age is the dominant parameter
which correlates with Hubble residuals, and that there is insufficient evidence
for any metallicity bias within SN~Ia studies.

\subsection{A SN~Ia sample in exclusively star-forming galaxies}
The currently analysed sample is formed exclusively by star-forming galaxies and lacks 
elliptical hosts. While the SN~Ia rate per unit mass is significantly higher in star-forming
galaxies (see e.g. \citealt{sul06}), ellipticals still produce a significant number 
of SNe~Ia. Distinct SN~Ia populations 
are found in ellipticals and in star-forming galaxies. The population within ellipticals is generally less luminous 
with lower stretch values. We have found that SN~Ia stretch does not appear to correlate with 
environment \textit{within} host galaxies. 
Given the classic result of a correlation between SN light-curve morphology with
host galaxy type --that higher stretch SNe are found in later
morphological types-- one may therefore ask how we reconcile this with our result
showing no environmental differences stretch separated samples.
It could be that our sample is biased in such
a way to remove those SNe driving the global stretch trend. To investigate we analyse host
galaxy T-types\footnote{Taken from the HyperLeda database: leda.univ-lyon1.fr.} for all SNe with estimated stretch values. In Fig.\ 16 we show the 
cumulative distributions of the SN~Ia sample with respect to host galaxy type, 
separated by median stretch. The classic result still holds, 
in that the high stretch sub-sample has in general host galaxies of
higher T-type. The difference between the two samples is significant at the 2.5\%\ level
when a KS-test is applied.
We also investigate host T-type distributions when the sample is split by the median
SN \textit{(B-V)}$_{max}$ colour, and they are presented in Fig.\ 17. This shows the possibly 
surprising result that `redder' SNe are found in earlier T-type hosts (significant at the $\sim$5\%\ level). 
If one assigns the majority of SN~Ia colour diversity to ISM
effects, then one would expect the opposite result. We note that a similar result
was presented by \cite{smi12}. This is possibly hinting at a metallicity effect, with `redder' events
occurring in higher metallicity (earlier Hubble type) galaxies.\\
\indent In conclusion, while overall global host galaxy properties (such as age and metallicity) appear 
to affect the light-curve shape of SNe~Ia, the specific stellar populations nearer
to the explosion sites of these events do not seem to affect their properties. This is in
contrast to CC SNe, where differences between global properties of host galaxies (see e.g. \citealt{hak14}) 
are in general
much smaller than differences of their nearby environments.
The main driver of these differing results would appear to 
be the ages of the respective progenitors. CC SNe have lifetimes of at most several tens of Myrs and 
hence they have little time to move away from their parent stellar populations. On the other hand, SNe~Ia
most likely have progenitor delay times of at least several 100 Myrs, which gives both their
progenitors time to move away from their birth sites, but also gives time for the population 
of stars at their birth sites to evolve significantly. This effectively washes out the information
that can be gained on SN~Ia progenitors from analysing the stellar populations at the exact explosion sites of SNe~Ia.

\begin{figure}
\includegraphics[width=8.5cm]{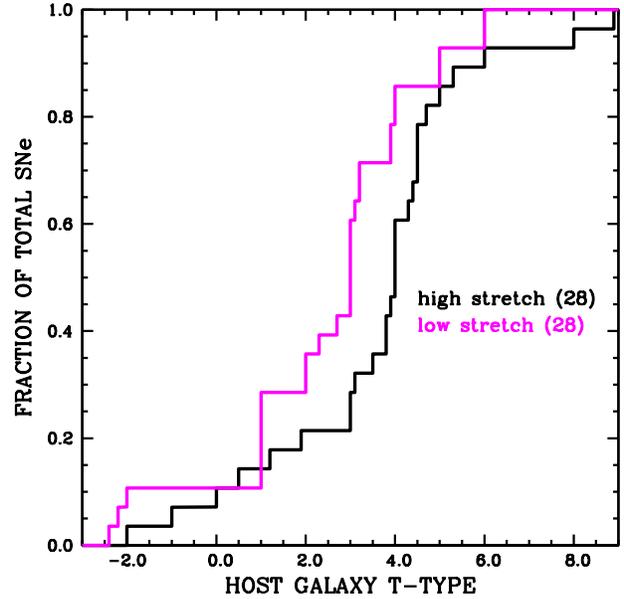}
\caption{Cumulative plot showing the host galaxy T-type distribution of 
SNe~Ia when divided into high and low stretch samples.}
\end{figure}

\begin{figure}
\includegraphics[width=8.5cm]{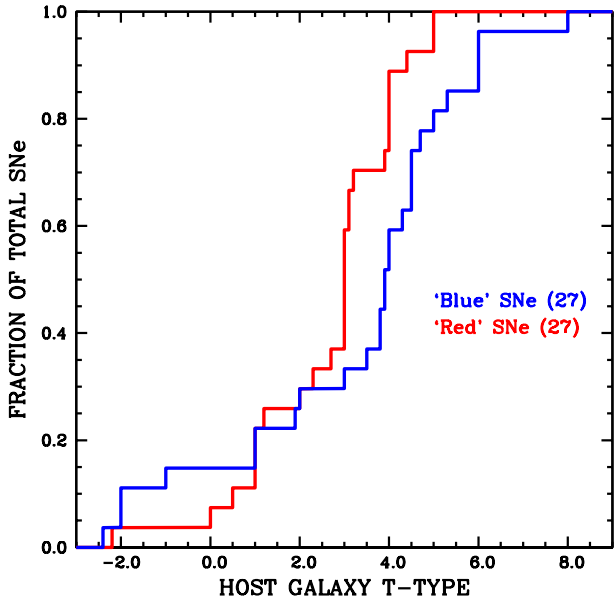}
\caption{Cumulative plot showing the host galaxy T-type distributions 
of SNe~Ia when divided into `red' and `blue' colour samples.}
\end{figure}

\subsection{SN colour with respect to environmental effects}

\begin{figure}
\includegraphics[width=9cm]{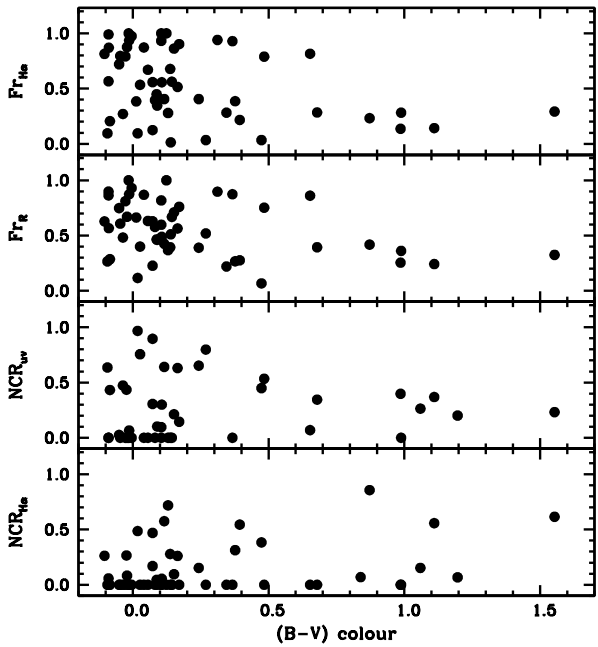}
\caption{Environmental property indicators against \textit{(B-V)}$_{max}$.}
\end{figure}

One of our key findings is that `redder' SNe are found to occur
both nearer to bright \hii\ regions, while at the same time closer to the centres of
hosts than their `bluer' counterparts. Given 
the results shown in \S\ 4.3, one may worry that it is the reddest events which are
driving these trends, where it is generally accepted that their `red' colours do indeed
arise from ISM extinction. To elucidate this issue we thus cut our sample to only include
those SNe with \textit{(B-V)}$_{max}$$<$0.2, which leaves one with a more general sample
similar to those usually used for cosmology.
Again the sample is split by the median \textit{(B-V)}$_{max}$. The above findings are 
completely robust to this further analysis. In this sub-set of
events the `redder' SNe are generally found to occur close to regions of \ha\ SF, and at the same time
within more central regions.\\
\indent In Fig.\ 18 we present SN \textit{(B-V)}$_{max}$ colours plotted against
our four main environment indicators: NCR$_{\textit{\ha}}$, NCR$_{nUV}$,
\textit{Fr}$_{\textit{R}}$, \textit{Fr}$_{\textit{\ha}}$.
The vast majority of SNe~Ia which have
\ha\ NCR values of zero are relatively `blue' events. There is also
a lack of `blue' SNe with NCR values higher than 0.4: all SNe
with negative \textit{(B-V)}$_{max}$ values have NCR statistics lower than this value,
however for `redder' events there are many SNe with higher values.
With respect to the UV NCR distribution, any trend with SN colour
is much less clear, as already shown above. The one obvious
observation is that again the majority of events with NCR values of zero are
clustered around \textit{(B-V)}$_{max}$ colours of zero. With respect to the 
radial values \textit{Fr}$_{\textit{R}}$ and \textit{Fr}$_{\textit{\ha}}$,
one can see that `redder' SN in general have much lower radial values. Indeed,
SNe with colours `redder' than 0.7 exclusively have radial values lower than 0.4
(and lower than 0.3 with respect to \textit{Fr}$_{\textit{\ha}}$). Meanwhile
`bluer' SNe show the full range of radial values, although 
the very bluest SNe with negative colours appear to have a preference
for exploding
in the outer regions of their galaxies, as indicated by their large \textit{Fr} values.\\
\indent In regions of more intense SF one expects a higher degree of extinction, which is also
expected for SNe occurring in more central regions. It is thus probable that 
the environments where these SNe explode are affecting their colours through line of sight ISM extinction effects,
i.e. material in the line of sight reddens the light emitted by the SNe through extinction.
A key property of SNe spectra which is assumed to trace the presence and amount of material 
within the line of sight is narrow sodium (NaD) absorption. The strength of this feature
is often used to constrain the amount of host galaxy extinction towards SNe (see e.g. discussion in
\citealt{poz12} and \citealt{phi13}). 
Hence, within environments where one expects a higher degree of extinction --closer to
\hii\ regions or within more central parts of galaxies-- one should also see higher
absorption due to line of sight material within SNe spectra. The equivalent width (EW) of the unresolved 
sodium doublet (NaD) was measured for all SNe with available spectra using the 
method outlined in \cite{for12} and further elaborated in \cite{for13}. We then split the samples
into those SNe with positive EW measurements, and those without. The results of this analysis 
are shown in in Figs\ 19 and 20, where we plot the \ha\ NCR and \textit{Fr}$_{\textit{R}}$ distributions
respectively for positive and zero EW measurement sub-samples. The results are striking.
With respect to \ha\ NCR, we find that the samples are statistically very different with there being 
less than a 0.1\%\ chance probability of the two EW distributions being drawn from the same parent population.
Indeed only 2 of of 20 (10\%) events with no EW detection fall on regions of positive \ha\ emission,
while this jumps to 24 out of 36 (67\%) for SNe with positive EW measurements.
Fig.\ 20 shows that SNe with NaD EW detections are also much more likely to be more central within their galaxies
than those without detections, with a $\sim$1\%\ chance of the two distributions being
drawn from the same parent population.\\
\indent The simplest way to explain these results is that ISM material is causing reddening of SNe
and also leading to NaD EW detections, where there is more ISM material in the line of sight
of SNe when 
they are found to be coincident with \hii\ regions and/or more centrally. (However, see \citealt{pan14} who
argue that colour diverstiy is not due to ISM extinction because of the non-correlation with host galaxy
extinction as implied from the Balmer decrement.) Circumstellar material
in the line of sight could also produce SNe which are both `redder' and are found to more commonly contain
NaD absorption. The presence of such material close to the SN could give clues to
the true nature of SN~Ia progenitors. This would then imply that environmental properties
of explosion sites of SNe~Ia are determined by a progenitor property and not simply a chance alignment of a
SN with ISM material. A problem with this interpretation is that while it is possible
that the radial distribution of SNe~Ia could be explained, it is not clear how the CSM material hypothesis 
can explain that SNe are more often found within \hii\ regions, given the large offset between the lifetimes
of \hii\ regions and those of SN~Ia progenitors (where above we have suggested the latter must be more than several
100 Myrs).
\textit{However,} there are
several observational results which further complicate this matter. It was first observed by \cite{ste11}
that absorption of NaD in SNe~Ia spectra show an excess of blue-shifted (with respect to host galaxy velocities) 
profiles. This was further confirmed by \cite{mag13}, who also found that those
events showing blue-shifted absorption also have stronger absorption components, and were 
generally found to occur within star-forming galaxies. It is very difficult to explain such blue-shifted profiles
through ISM properties. In addition, both NaD absorption properties \citep{for12,for13},
and SN colours (see e.g. \citealt{fol11b,fol12b,mae11})
have been shown to correlate with many other SN~Ia properties, arguing that at least 
a fraction of their diversity is intrinsic to the SNe themselves (and not solely from ISM effects).
We also note that the centralisation of `red' SNe~Ia could hint at a metallicity effect. This would
then be consistent with the global host properties of our sample when split by SN colour, as presented in Fig.\ 17.\\
\indent Returning to Fig.\ 19 there is another intriguing observations that
has yet to be mentioned. Around 25\%\ of SNe~Ia \textit{with} NaD absorption detections fall
on regions of \textit{zero \ha\ flux}. It is possible that this is simply
evidence of significant host galaxy extinction outside of bright \hii\ regions.
However, an exciting possibility is that these SNe are those where the NaD features are
produced by CSM material. Future environment studies correlating the presence of blue-shifted
Na D absorption with host galaxy SF may be particularly revealing in this context.\\
\indent It is clear that distinguishing between the effects of ISM and CSM extinction for 
SNe~Ia could have profound implications for the discussion of their progenitors. However, as the above shows,
such a differentiation is observationally difficult. We have presented further evidence that line of sight 
material plays a significant role in determining some of the transient features of SNe~Ia. Further work will
be needed to further elucidate what this means for SN~Ia progenitors.

\begin{figure}
\includegraphics[width=8.5cm]{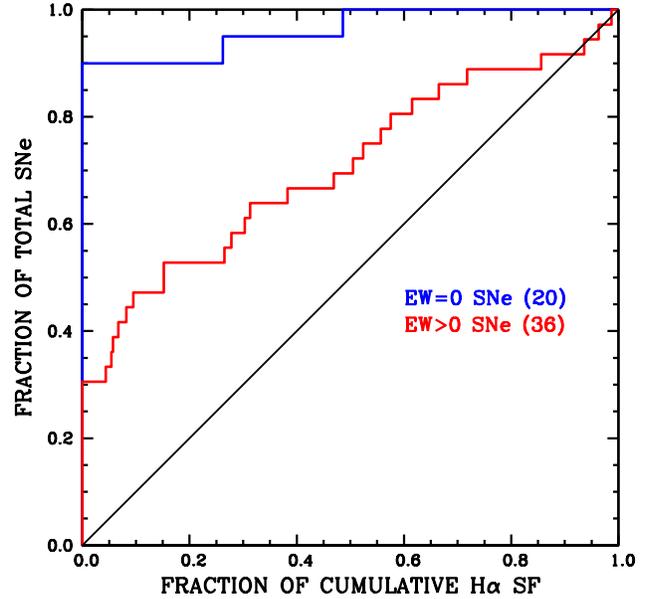}
\caption{Cumulative \ha\ NCR plot showing distributions of SNe~Ia when split into those
SNe with and those SNe without NaD absorption in their spectra.}
\end{figure}

\begin{figure}
\includegraphics[width=8.5cm]{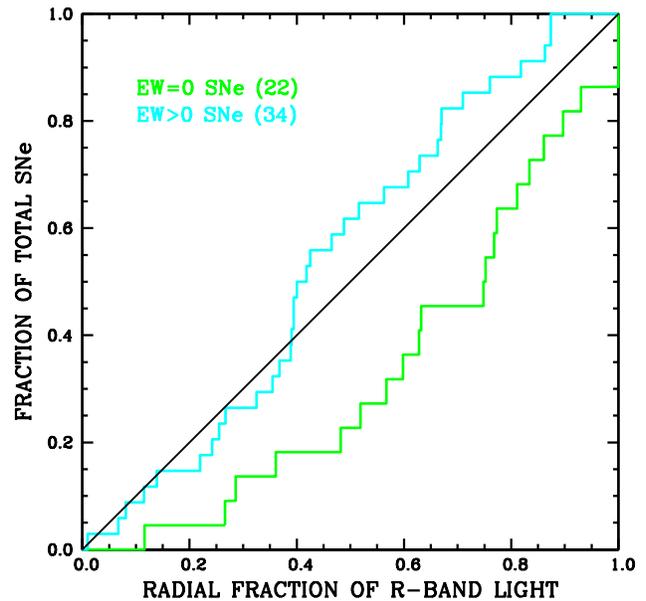}
\caption{Cumulative plot showing the \textit{Fr}$_{\textit{R}}$ distributions of SNe~Ia
when the sample is split into those SNe with and those SNe without NaD absorption in their spectra.}
\end{figure}

\subsection{In the context of previous work}
A first study
of the immediate environments of SNe~Ia was presented by \cite{rig13}. These authors measured
or set limits on \ha\ emission at the locations of 89 SNe~Ia, where they included
elliptical host galaxies. Similar to our work, \citeauthor{rig13} showed that moderate stretch 
SNe were found in all types of \ha\ environment. With respect to colour
they observed that `redder' SNe were found in higher star-forming environments with a higher flux of \ha\ emission. 
This
is equivalent to our finding using \ha\ NCR pixel statistics (see \S\ 4.1.4 and Fig.\ 5). Unfortunately
we are not in a position to test for environmental effects on Hubble residuals in our sample, as 
done by those authors, due 
to the low number of events which could be included in such an analysis.
\cite{wan13} investigated how spectral velocities of SNe~Ia correlate with local environment
in terms of radial distributions of SNe (but using normalised distances in place
of normalising to flux as done here) and pixel statistics with respect to $u, g, r$-band light 
(following the formalism of \citealt{fru06}).
Similar to our findings, they found that overall the SN~Ia population followed the $g$ and $r$ (somewhat equivalent to 
the analysis presented with respect to the $B$ and $R$ light distributions) reasonably well, but not
the $u$-band light (see their Fig.\ 3). \citeauthor{wan13} in particular separated the SN~Ia sample into
high and low velocity events, claiming for two distinct populations given the differences 
in environment of the two sub-samples.
High velocity events were found to be more centrally within hosts and also to explode on brighter regions
(although note the lower significance of this trend in a separate sample as shown by \citealt{pan15}).
While these results are intriguing, we caution on mixing spiral and elliptical hosts when analysing environments. The stellar 
populations and their distributions are extremely distinct, and it makes more sense to analyse the two samples
separately. Here we do not include such an analysis of environments with respect to spectral velocities
within exclusively star-forming host galaxies, but 
such a study in the future may be revealing to confirm the above claims.\\
\indent The radial distribution of SNe~Ia was found to show a deficit in the central regions of spiral galaxies
by \cite{ber97}, as observed above (also see \citealt{wan97}). \cite{gal12} analysed the radial distribution of a sample
$\sim$200 SNe~Ia and concluded that SNe in the more central parts of spiral host galaxies SNe~Ia have 
larger colours (or equivalent parameters)
than those SNe in the outer regions, similar to what we have shown. 
A significant difference between our study and those previously
is that we normalise radial positions to flux and not distance, meaning we are analysing where within the radial
light-distribution SNe fall.\\
\indent Overall the results we present are consistent with those previously presented in the literature.
SN stretch does not show any significant correlation with local environment in star-forming galaxies, seen through either
measurements of the flux at the exact explosion sites, or through analyses of the radial distributions of stretch sub-samples.
However, SN colour does seem to show significant correlation with environment.

\section{Conclusions}
We have presented statistics on the properties of the environments 
of SNe~Ia within star-forming galaxies. Our main findings are that
SNe~Ia best trace the $B$-band light distribution within their galaxies, while
they do not trace neither the extreme young populations traced by \ha\ or near-UV emission,
nor the evolved populations traced by $J$- and $K$-band light.
It is found that `redder' SNe are more often found both to be coincident with \hii\ regions and
are more centrally concentrated within their host galaxies. In addition,
SNe with positive NaD EW measurements
are also much more likely to be found in central regions and coincident with bright \hii\ regions. Whether this effect
is dominated by ISM or CSM material, or some other progenitor property,
remains unresolved.
We find a central deficit of SNe~Ia with respect to the radial
distribution of stellar continuum.
Finally, the main conclusions to arise from this investigation are listed below.
\begin{itemize}
\item
The SN~Ia population best traces the $B$-band light of their host galaxies.
This is in contrast to the \ha, near-UV, $J$- and $K$-band light, where the SNe are inconsistent
with being drawn from the underlying populations traced by those bands.
\item
`Redder' SNe are found to occur closer to \hii\ regions and also more centrally than their `blue' counterparts. 
This implies that a significant source of 
SN colour arises from line of sight material producing extinction and hence reddening effects.
\item
While we recover the host galaxy type stretch relation found by numerous previous authors, we find no
evidence that stretch is related to environments \textit{within} host galaxies.
\item
SNe~Ia within star-forming galaxies do not trace the underlying
on-going (\ha) or recent (near-UV) SF. This implies that the dominant
population of
SNe~Ia in star-forming galaxies do not originate from progenitors with delay times of
less than a few 100 Myr.
\item
Brighter SNe~Ia within host galaxies do not follow the spatial distribution of
recent SF, implying that even even those SNe~Ia associated with a relatively
young stellar population do not arise from extreme `prompt' progenitor
channels. This effectively rules out progenitor delay times of less than a few
100 Myr for all but a small minority of events
\item
A deficit of SNe~Ia is found within the central 20\%\ of the $R$-band stellar continuum. 
There is also a suggestion that more SNe~Ia per unit SF are produced in the outer
regions of galaxies. This observation could be a metallicity effect with lower metallicity
populations producing a higher fraction of SNe. Or it could be explained through
a progenitor age effect where the central regions of these galaxies have SFHs
weighted to older ages, and hence more SNe are produced in the outer regions
where younger stellar populations dominate.
\end{itemize}

\section*{Acknowledgments}
We thank the referee Mike Childress for his positive report, and useful suggestions on
improving the manuscript.
Mark Sullivan, Gaston Folatelli, Giuliano Pignata, Mark Phillips, Lluis Galbany, Mike Bode and Stephen Smartt are 
thanked for useful 
discussion.
We thank the Carnegie Supernova Project for allowing us to use light-curve fits
to unpublished photometry of SN~2009ag.
M.~H., F.~F., and S.~G. acknowledge support from the Millennium Institute of 
Astrophysics (MAS; Programa Iniciativa Científica Milenio del Ministerio 
de Economía, Fomento y Turismo de Chile, grant IC120009).
S.~G. thanks CONICYT through FONDECYT grant 3110142.
The Liverpool Telescope is operated on the island of La Palma by Liverpool 
John Moores University in the Spanish Observatorio del Roque de los Muchachos 
of the Instituto de Astrofisica de Canarias with financial support from the UK 
Science and Technology Facilities Council. 
Based 
on observations made with the Isaac Newton Telescope
operated on the island
of La Palma by the Isaac Newton Group in the Spanish Observatorio del Roque de los
Muchachos of the Instituto de Astrofisica de Canarias,
and observations made with the 2.2m MPG/ESO telescope at La Silla, proposal
ID: 084.D-0195 and based on observations made with the Nordic Optical Telescope, 
operated by the Nordic Optical Telescope Scientific Association at the Observatorio 
del Roque de los Muchachos, La Palma, Spain, of the Instituto de Astrofisica de Canarias
This research
has made use of the NASA/IPAC Extragalactic Database (NED) 
which is operated by the Jet 
Propulsion Laboratory, California
Institute of Technology, under contract with the National Aeronautics.
Some of the data presented in this paper were obtained from the 
Mikulski Archive for Space Telescopes (MAST). STScI is operated by the 
Association of Universities for Research in Astronomy, Inc., under 
NASA contract NAS5-26555. Support for MAST for non-HST data is provided 
by the NASA Office of Space Science via grant NNX13AC07G and by other grants and contracts.

\bibliographystyle{mn2e}

\bibliography{Reference}

\appendix

\section[]{SN environment statistics}

\begin{table*}\centering
\caption{SNe environment statistical analyses values. In the first column the SN name is listed. 
This is followed by the NCR values derived from \ha, near-UV, $B$-band, $R$-band, 
$J$-band and $K$-band in columns 2, 3, 4, 5, 6, and 7 respectively. In columns 8 and 9 the 
\textit{Fr}$_{\textit{R}}$ and \textit{Fr}$_{\textit{\ha}}$ are listed.}
\begin{tabular}[t]{cccccccccc}
\hline
\hline
SN & NCR$_{\textit{\ha}}$ & NCR$_{nUV}$ &NCR$_{B}$ & NCR$_{R}$&NCR$_{J}$ & NCR$_{K}$&
\textit{Fr}$_{\textit{R}}$ & \textit{Fr}$_{\textit{\ha}}$\\
\hline
1937C	&0.555&	0.240&	$\cdots$&$\cdots$	&$\cdots$&$\cdots$&0.272&0.340\\
1954B	&0.466&	0.537&	0.629	&0.719	&0.440	&0.000  &0.253&0.137\\
1957A	&0.000&	0.572&	$\cdots$&0.076	& $\cdots$& $\cdots$&0.797&0.831\\
1963I	&0.317&	0.319&	0.792	&0.557	&0.625	&0.808  &0.111&0.068\\
1963J	&0.000&	$\cdots$&	0.721	&0.346	&0.466	&0.000  &0.307&0.126\\
1968E	&0.277&	0.599&	0.567	&0.503	&0.485	&0.471  &0.560&0.577\\
1968I	&0.000&	0.000&	0.639	&0.822	&0.769	&0.774  &0.137&0.063\\
1969C	&0.297&	$\cdots$&$\cdots$&0.709	& $\cdots$&$\cdots$&0.300&0.178\\
1971G	&0.030&	0.038&$\cdots$&0.011	&0.422	&0.000  &0.794&0.956\\
1972H	&0.000&	0.281&	0.498	&0.344	&0.253	&0.016  &0.646&0.750\\
1974G	&0.000&	0.367&	0.000	&0.336	&0.137	&0.032  &0.772&0.816\\
1975A	&0.000&	0.000&	$\cdots$&$\cdots$&$\cdots$&$\cdots$&0.722&0.672\\
1979B	&0.000&	$\cdots$&	0.288	&0.725	&0.000	&0.000  &0.784&0.670\\
1981B	&0.000&	$\cdots$&$\cdots$&0.158	&$\cdots$&$\cdots$&0.731&0.747\\
1982B	&0.000&	0.484&	0.560	&0.367	&0.453	&0.278  &0.507&0.572\\
1983U	&0.000&	$\cdots$&$\cdots$&$\cdots$&$\cdots$&$\cdots$&0.365&0.735\\
1986A	&0.249&	0.038&	0.735	&0.682	&0.397	&0.000  &0.433&0.437\\
1986G	&0.069&	$\cdots$&$\cdots$&$\cdots$&$\cdots$&$\cdots$&$\cdots$&$\cdots$\\
1987D	&0.000&	0.356&$\cdots$&0.425	&$\cdots$&$\cdots$&0.485&0.963\\
1987O	&0.000&	0.000&$\cdots$&$\cdots$&$\cdots$&$\cdots$&0.672&0.755\\
1989A	&0.000&	0.238&	$\cdots$		&0.160	&0.000	&0.000  &0.674&0.726\\
1989B	&0.544&	$\cdots$&$\cdots$&0.732	&$\cdots$&$\cdots$&0.276&0.216\\
1990N	&0.000&	0.000&$\cdots$&0.089	&$\cdots$&$\cdots$&0.869&0.872\\
1991ak	&0.000&	0.652&	0.454	&0.363	&0.301	&0.000  &0.645&0.655\\
1991T	&0.000&	0.000&	0.355	&0.072	&0.126	&0.253  &0.578&0.395\\
1992bc	&0.000&	0.000&$\cdots$&0.000	&$\cdots$&$\cdots$&0.898&0.566\\
1992G	&0.343&	0.539&	0.757	&0.761	&0.707	&0.630  &0.335&0.351\\
1992K	&0.395&	0.663&	$\cdots$&$\cdots$&$\cdots$&$\cdots$&0.214&0.156\\
1994ae	&0.000&	0.026&	0.261	&0.205	&0.000	&0.099  &0.748&0.720\\
1994S	&0.082&	0.000&	$\cdots$&0.278	&$\cdots$&$\cdots$&0.670&0.876\\
1995al	&0.054&	0.300&	0.333	&0.328	&0.330	&0.234  &0.488&0.557\\
1995D	&0.000&	0.000&	0.000	&0.000	&0.000	&0.000  &0.930&0.975\\
1995E	&0.000&	0.345&	0.881	&0.862	&0.846	&0.737  &0.394&0.284\\
1996ai	&0.615&	0.232&	0.810	&0.801	&0.661	&0.661  &0.325&0.292\\
1996Z	&0.000&	$\cdots$&$\cdots$&0.000	&$\cdots$&$\cdots$&0.768&0.748\\
1997bp	&0.095&	0.213&$\cdots$&0.266	&$\cdots$&$\cdots$&0.710&0.862\\
1997bq	&0.000&	0.096&	0.184	&0.170	&0.000	&0.000  &0.818&0.932\\
1997do	&0.486&	0.967&	0.940	&0.905	&0.749	&0.736  &0.116&0.095\\
1997dt	&0.524&	$\cdots$&$\cdots$&0.916	&$\cdots$&$\cdots$&0.355&0.281\\
1997Y	&0.303&	$\cdots$&	0.371	&0.810	&0.633	&0.467  &0.139&0.259\\
1998aq	&0.262&	$\cdots$&	0.460	&0.426	&0.313	&0.083  &0.628&0.814\\
1998bu	&0.000&	0.798&	$\cdots$&0.424	&$\cdots$&$\cdots$&0.519&0.034\\
1998D	&0.000&	0.000&	0.337	&0.178	&0.208	&0.164  &0.608&0.795\\
1998dh	&0.044&	$\cdots$&$\cdots$&0.339	&$\cdots$&$\cdots$&0.465&0.449\\
1998eb	&0.000&	0.175&$\cdots$&$\cdots$&$\cdots$&$\cdots$&0.516&0.554\\
1999aa	&0.000&	0.473&	0.638	&0.600	&0.398	&0.513  &0.482&0.270\\
1999bh	&0.856&	$\cdots$&	0.599	&0.446	&0.348	&0.450  &0.418&0.232\\
1999bv 	&0.000&	$\cdots$&$\cdots$&0.000	&$\cdots$&$\cdots$&0.834&0.618\\
1999by	&0.000	&0.534	&$\cdots$&0.160	&$\cdots$&$\cdots$&0.752&0.789\\
1999cl	&0.152	&0.264	&$\cdots$&$\cdots$&$\cdots$&$\cdots$&$\cdots$&$\cdots$\\
1999cp	&0.000	&0.000	&0.000	&0.073	&0.000	&0.000  &0.811&0.791\\
1999gd	&0.000	&0.000	&0.359	&0.346	&0.186	&0.000  &0.874&0.929\\
2000ce	&0.505	&$\cdots$&0.500	&0.392	&0.000	&0.135  &0.563&0.507\\
2000E	&0.718	&0.000	&$\cdots$&$\cdots$&$\cdots$&$\cdots$&0.368&0.279\\
2001ay	&$\cdots$&$\cdots$&$\cdots$&$\cdots$&$\cdots$&$\cdots$&0.638&0.384\\
2001bg	&0.000	&0.145	&0.207	&0.212	&0.000	&0.000  &0.760&0.902\\
2001cz	&0.000	&0.102	&$\cdots$&0.363	&$\cdots$&$\cdots$&0.461&0.346\\
2001E	&0.265	&0.435	&0.503	&0.477	&0.283	&0.133  &0.663&0.384\\
2001eg	&0.000	&0.000	&0.225	&0.265	&0.000	&0.606  &0.605&0.774\\
2002au	&0.096	&0.825	&0.481	&0.374	&0.344	&0.000  &0.764&0.864\\
2002bs	&0.963	&0.988	&0.993	&0.991	&0.956	&0.934  &0.010&0.009\\
2002cr	&0.000	&0.066	&0.000	&0.052	&0.000	&0.000  &0.874&0.936\\
2002er	&0.278	&0.000	&$\cdots$&0.564	&0.180  &$\cdots$&0.394&0.678\\
\hline	       
\end{tabular}
\end{table*}

\setcounter{table}{0}

\begin{table*}\centering
\caption{\textit{Continued...}}
\begin{tabular}[t]{ccccccccccc}
\hline
\hline
SN & NCR$_{\textit{\ha}}$ & NCR$_{nUV}$ &NCR$_{B}$ &NCR$_{R}$ &NCR$_{J}$ & NCR$_{K}$&
\textit{Fr}$_{\textit{R}}$ & \textit{Fr}$_{\textit{\ha}}$\\
\hline
2002fk	&0.000	&0.636	&$\cdots$&$\cdots$&$\cdots$&$\cdots$&0.266&0.095\\
2003cg	&0.557	&0.369	&$\cdots$&0.693	&$\cdots$&$\cdots$&0.242&0.142\\
2003cp	&0.000	&$\cdots$&0.606	&0.509	&0.023	&0.000  &0.494&0.605\\
2003du	&0.000	&0.432	&0.670	&0.000	&0.000	&$\cdots$&0.286&0.205\\
2004bc	&0.987	&0.821	&$\cdots$&0.966	&$\cdots$&$\cdots$&0.389&0.189\\
2004bd	&0.665	&0.214	&0.865	&0.862	&0.759	&0.769  &0.115&0.434\\
2005A	&0.000	&0.398	&$\cdots$&0.506	&$\cdots$&$\cdots$&0.255&0.136\\
2005am	&0.000	&0.000	&0.330	&0.000	&0.265	&0.255  &0.632&0.670\\
2005bc	&0.313	&$\cdots$		&0.770	&0.751	&0.716	&0.725  &0.267&0.386\\
2005bo	&0.152	&0.653	&0.387	&0.456	&0.420	&0.376  &0.390&0.405\\
2005cf	&0.000	&0.000	&$\cdots$&$\cdots$&$\cdots$&$\cdots$&1.000&1.000\\
2005el	&0.000	&$\cdots$&$\cdots$&0.000	&$\cdots$&$\cdots$       &0.567&0.870\\
2005F	&$\cdots$&$\cdots$&$\cdots$&$\cdots$&$\cdots$&$\cdots$&0.975&1.000\\
2005G	&$\cdots$&$\cdots$&$\cdots$&$\cdots$&$\cdots$&$\cdots$&0.773&0.656\\
2005ke	&0.000	&0.069	&$\cdots$&$\cdots$		& $\cdots$      & $\cdots$      &0.861&0.815\\
2005M	&$\cdots$&$\cdots$&$\cdots$&$\cdots$		& $\cdots$		& $\cdots$		&0.897&0.940\\
2005W	&0.000	&0.000	&$\cdots$&$\cdots$		& $\cdots$      & $\cdots$      &0.669&0.562\\
2006ax	&0.057	&0.000	&$\cdots$		&0.057	& $\cdots$      & $\cdots$      &0.863&0.990\\
2006ce	&0.000	&0.000	&$\cdots$		&0.090	& $\cdots$      & $\cdots$      &0.850&0.887\\
2006D	&0.000	&0.000	&$\cdots$		&0.366	& $\cdots$      & $\cdots$      &0.598&1.000\\
2006mq	&0.000	&$\cdots$&$\cdots$&$\cdots$&$\cdots$&$\cdots$&$\cdots$&$\cdots$\\
2006N	&0.000	&0.756	&0.560	&0.591	&0.451	&0.592  &0.400&0.534\\
2006ou	&0.000	&$\cdots$		&$\cdots$	    &0.562	&$\cdots$       &$\cdots$       &0.220&0.282\\
2006X	&0.067	&0.201	&$\cdots$		&$\cdots$		&$\cdots$       &$\cdots$       &$\cdots$&$\cdots$\\
2007af 	&0.469	&0.306	&$\cdots$		&0.372	&$\cdots$       &$\cdots$       &0.629&0.558\\
2007bm	&0.383	&0.449	&$\cdots$		&0.910	&$\cdots$       &$\cdots$       &0.067&0.034\\
2007N	&0.000	&0.000	&$\cdots$		&0.429	&$\cdots$       &$\cdots$       &0.361&0.282\\
2007S	&0.575	&0.641	&$\cdots$		&0.526	&$\cdots$       &$\cdots$       &0.425&0.405\\
2007sr	&0.000	&$\cdots$&$\cdots$&$\cdots$&$\cdots$&$\cdots$&1.000&1.000\\
2008bi	&0.936	&0.722	&$\cdots$&0.906	&$\cdots$&$\cdots$&0.081&0.029\\
2008fv	&0.261	&0.631	&0.373	&0.478	&0.318	&0.152  &0.564&0.515\\
2009ag	&0.000	&$\cdots$		&$\cdots$	    &$\cdots$&$\cdots$&$\cdots$&1.000&1.000\\
2009ds	&0.169&	0.896&	0.771	&0.767	&0.722	&0.909	&0.227&0.124\\
2009ig	&0.000&	$\cdots$&$\cdots$  &0.480	&$\cdots$       &$\cdots$     	&0.511&0.013\\
2010eb	&0.000&	0.000&$\cdots$&0.000	&$\cdots$       &$\cdots$     	&1.000&1.000\\
2011ao	&0.000&	0.301&$\cdots$&0.545	&$\cdots$       &$\cdots$     	&0.398&0.379\\
2011B	&$\cdots$&$\cdots$&$\cdots$&		&$\cdots$		&$\cdots$		&0.247&1.000\\
2011dm	&0.000	&$\cdots$&$\cdots$&$\cdots$&$\cdots$       &$\cdots$     	&1.000&1.000\\
2011dx 	&0.000&	0.209&$\cdots$&0.218	&$\cdots$       &$\cdots$     	&0.755&0.706\\
2011ek	&0.000&	0.000&$\cdots$&0.000	&$\cdots$&$\cdots$     	&1.000&1.000\\
\hline    
\end{tabular}
\setcounter{table}{0} 
\end{table*}

\label{lastpage}

\end{document}